\documentclass[12pt]{article}

\usepackage[utf8]{inputenc}
\usepackage[T1]{fontenc}

\usepackage[compress]{cite}
\usepackage{IEEEtrantools,stackengine}
\stackMath

\usepackage{authblk}
\usepackage{fullpage}
\usepackage{amssymb,amsmath}
\usepackage{adjustbox}
\usepackage{graphicx}
\usepackage{longtable}

\usepackage[most]{tcolorbox}
\definecolor{darkheader}{RGB}{60, 60, 60}  
\definecolor{lightbody}{RGB}{250, 250, 250} 

\usepackage{siunitx}
\usepackage{csquotes}

\usepackage{xspace}

\usepackage{booktabs}
\usepackage{longtable}
\usepackage{multirow}

\usepackage[dvipsnames, table]{xcolor}

\definecolor{darkgreen}{rgb}{0.0, 0.5, 0.0}

\definecolor{lightyellow}{HTML}{FFE699}
\definecolor{red_revision}{HTML}{FF0000}

\usepackage{setspace}
\usepackage[unicode=true]{hyperref}
\hypersetup{breaklinks=true,
            pdfborder={0 0 0},
            allcolors=black}
\usepackage[%
  sort&compress
]{cleveref}
  \Crefname{appendix}{Supplement}{Supplements}
  \Crefname{figure}{Fig.}{Fig.}

\usepackage{times}
\usepackage{enumerate}
\usepackage{enumitem}
\usepackage{float}

\usepackage{subcaption}
\usepackage{graphicx}

\usepackage{rotating}
\usepackage{tabularx}
\usepackage{dcolumn}
\usepackage{pdflscape}
\usepackage{rotating}

\usepackage{array}

\usepackage{titlesec}

\titleformat{\subsubsection}
  {\normalfont\itshape}{\thesubsubsection}{1em}{}

\titleformat{\subsection}
  {\normalfont\bfseries}{\thesubsection}{1em}{}

\makeatletter
\renewcommand{\fps@figure}{H}         
\renewcommand{\fps@table}{H}         
\makeatother 

\renewcommand{\arraystretch}{1.2}

\allowdisplaybreaks

\usepackage{eurosym}
\usepackage{comment}

\usepackage{ntheorem}
\theoremseparator{:}

\usepackage{makecell}

\title{\centering\LARGE\singlespacing A meta-analysis of the effect of generative AI on productivity and learning in programming}

\author[1,2,$\dagger$]{Sebastian Maier}
\author[1,$\dagger$]{Moritz Gunzenhäuser}
\author[1,2]{Jonas Schweisthal}
\author[1]{Manuel Schneider}
\author[1,2]{Stefan Feuerriegel\thanks{Corresponding author: feuerriegel@lmu.de}}

\affil[1]{LMU Munich, Munich, Germany}
\affil[2]{Munich Center for Machine Learning (MCML), Munich, Germany}
\affil[$\dagger$]{These authors contributed equally to this work.}

\date{}

\begin{document}

\maketitle
\newpage

\begin{abstract}\normalfont
\noindent
Generative artificial intelligence (GenAI) is increasingly used for programming, yet it remains unclear when and where GenAI tools lead to productivity gains. Evidence on the effects of GenAI on the long-term development of programming skills is similarly mixed. Here, we present a meta-analysis of $n = 23$ studies reporting $k = 27$ effect sizes to quantify the effect of GenAI-powered coding assistants on productivity and learning. We systematically searched (i)~ACM, (ii)~arXiv, ~(iii)~Scopus, and (iv)~Web of Science for studies published between 2019 and 2025. Studies were required to compare GenAI-assisted with unassisted programming using quantitative measures of (1) productivity (i.e., task completion time, commits, and lines of code) and (2) learning (i.e., exam performance). We assessed the risk of bias using RoB2 and ROBINS-I and compared standardized effect sizes using Hedges' $g$. We find a statistically significant, but moderate positive effect of GenAI assistance on developer productivity ($g = 0.33$, $95\%$ CI: $[0.09, 0.58]$), yet with substantial heterogeneity across settings. Notably, productivity gains tend to be larger in controlled experimental settings, while effects are smaller in open-source and enterprise contexts. In contrast, we find no statistically significant effect of GenAI assistance on learning outcomes ($g = 0.14$, $95\%$ CI: $[-0.18, 0.47]$). Overall, these results highlight that GenAI coding assistants can increase developer productivity, although these gains depend strongly on context. In educational settings, however, the use of GenAI does not consistently translate into improved learning or skill development, which highlights the need for careful integration of GenAI into computer science education.
\end{abstract}

\flushbottom
\maketitle
\thispagestyle{empty}

\sloppy
\raggedbottom


\newpage
\section*{Introduction}
\label{sec:introduction}


Generative artificial intelligence (GenAI) tools such as \textit{GitHub Copilot} \cite{GithubCopilot2025} and \textit{Cursor} \cite{noauthor_cursor_2025} have rapidly become a crucial part of the software engineering landscape \cite{banh_copiloting_2025, russo_manifesto_2024, feuerriegel2024generative}. Adoption rates of GenAI tools for coding grow rapidly \cite{daniottiWhoUsingAI2026}, with  84\% of developers already using or planning to use GenAI tools in their workflow \cite{stack_overflow_stack_2025}. This leads to a fundamental paradigm shift in how code is written and maintained \cite{qiu_todays_2024, ulfsnes_transforming_2024, otten_prompting_2025}. Many GenAI tools based on state-of-the-art large language models, such as \textit{GPT}, \textit{Claude}, \textit{Gemini}, and \textit{DeepSeek}, now achieve near-perfect scores on established coding benchmarks and reach or even surpass the vast majority of human developers \cite{jain2024livecodebench, DeepSeek_v3.2_2025, noauthor_gemini_nodate, quan2025codeelo, chen2021evaluating}. At the same time, GenAI has enabled ``vibe coding'', a style of development in which programmers iteratively interact with GenAI through prompts and accept generated code based on perceived plausibility rather than a detailed understanding of its logic or behavior \cite{Ge_VibeCoding2025, pimenova_good_2025, fawzy_vibe_2025}. Here, we analyze the effect of GenAI tools on productivity and learning in programming.


Empirical evidence on the productivity effects of GenAI tools remains mixed. On the one hand, developers report productivity gains from automating repetitive coding tasks and from lower cognitive load required for recalling syntax \cite{russo_navigating_2024, weisz_examining_2025, lyu_my_2025 ,marchesi_copilots_2025}. GenAI tools can also lower entry barriers and help developers, particularly novice developers, to understand unfamiliar code bases \cite{peng_impact_2023, prather_widening_2024, nam_using_2024}. On the other hand, using GenAI for programming changes work from active problem solving to passive evaluation and review of AI-generated code \cite{barkeGroundedCopilotHow2023}. This shift may also increase cognitive load and thus decrease performance, especially for experienced developers \cite{banh_copiloting_2025, Xu_aiassisteddecrease2025, becker_measuring_2025, liangLargescaleSurveyUsability2024}. In addition, AI-generated code can introduce errors such as security vulnerabilities \cite{perry2023users, pearce2025asleep} and ``technical debt'' due to reduced code maintainability and deviations from established design practices\cite{yetistiren_assessing_2022}. Vibe coding further encourages software development based on plausibility rather than true understanding of the underlying code \cite{Ge_VibeCoding2025, pimenova_good_2025, fawzy_vibe_2025}. While this may speed up initial development, it can increase maintenance costs over time, particularly for ensuring code quality \cite{Xu_aiassisteddecrease2025}. In sum, the impact of GenAI on productivity can vary widely across developers, tasks, and settings.


GenAI use is also reshaping how programming skills are acquired. Although proficiency with GenAI tools is increasingly expected in the labor market, such use by novice learners without well-developed foundational skills may alter how students engage with core learning processes and thus raises important challenges for computer science education \cite{kirova_software_2024, lau_ban_2023}. On the one hand, students benefit from personalized AI-based tutoring, for example, by asking GenAI systems to explain code fragments or solution strategies, which provides on-demand answers and explanations that are perceived as helpful and accessible \cite{aviv_impact2024, leinonen_comparing_2023, prather_widening_2024, laato_ai-assisted_2023}. On the other hand, reliance on GenAI can allow students to bypass core learning processes, such as trial-and-error learning, debugging, and iterative refinement, when they delegate core tasks to GenAI tools \cite{becker_programming_2023, guner_ai_2025, lee_impact_2025}. This phenomenon is often referred to as cognitive offloading \cite{vivian_cognitiveoffload_2025, jose_cognitive_2025} and may reduce active engagement with problem solving, which, in turn, can lead to shallow understanding and ultimately impair the development of fundamental programming skills \cite{rahe_how_2025, bastaniGenerativeAIGuardrails2025c}. As a result, the capabilities students develop during AI-assisted coding may prove tool-dependent, thus failing to persist without AI support \cite{wiles2024genai} or even yielding lower skill acquisition compared to unassisted learners \cite{shenHowAIImpacts2026a}.


Despite a growing body of work on GenAI-assisted programming, evidence on the effects on productivity and learning remains fragmented and often inconsistent \cite{becker_measuring_2025, cui_effects_2024, Xu_aiassisteddecrease2025, paradis_how_2024}. A key challenge is the substantial heterogeneity across studies, including differences in developer experience, and study settings (e.g., controlled experiments with standardized tasks versus real-world contexts). This heterogeneity makes it difficult to draw generalizable conclusions and leaves the question of when and where GenAI tools are effective. To address this, we focus on two core outcomes of GenAI-assisted programming: (1)~\emph{productivity}, operationalized through task completion time and code output, and (2)~\emph{learning}, measured by exam performance. We thus study the following two research questions (RQs): \textit{RQ1: How does GenAI assistance affect programmer productivity?} \textit{RQ2: How does GenAI assistance affect learning outcomes for computer science students?}

\begin{figure}
\centering
\includegraphics[
        width=\linewidth,
        trim=0cm 0cm 0cm 0cm, 
        clip
        ]{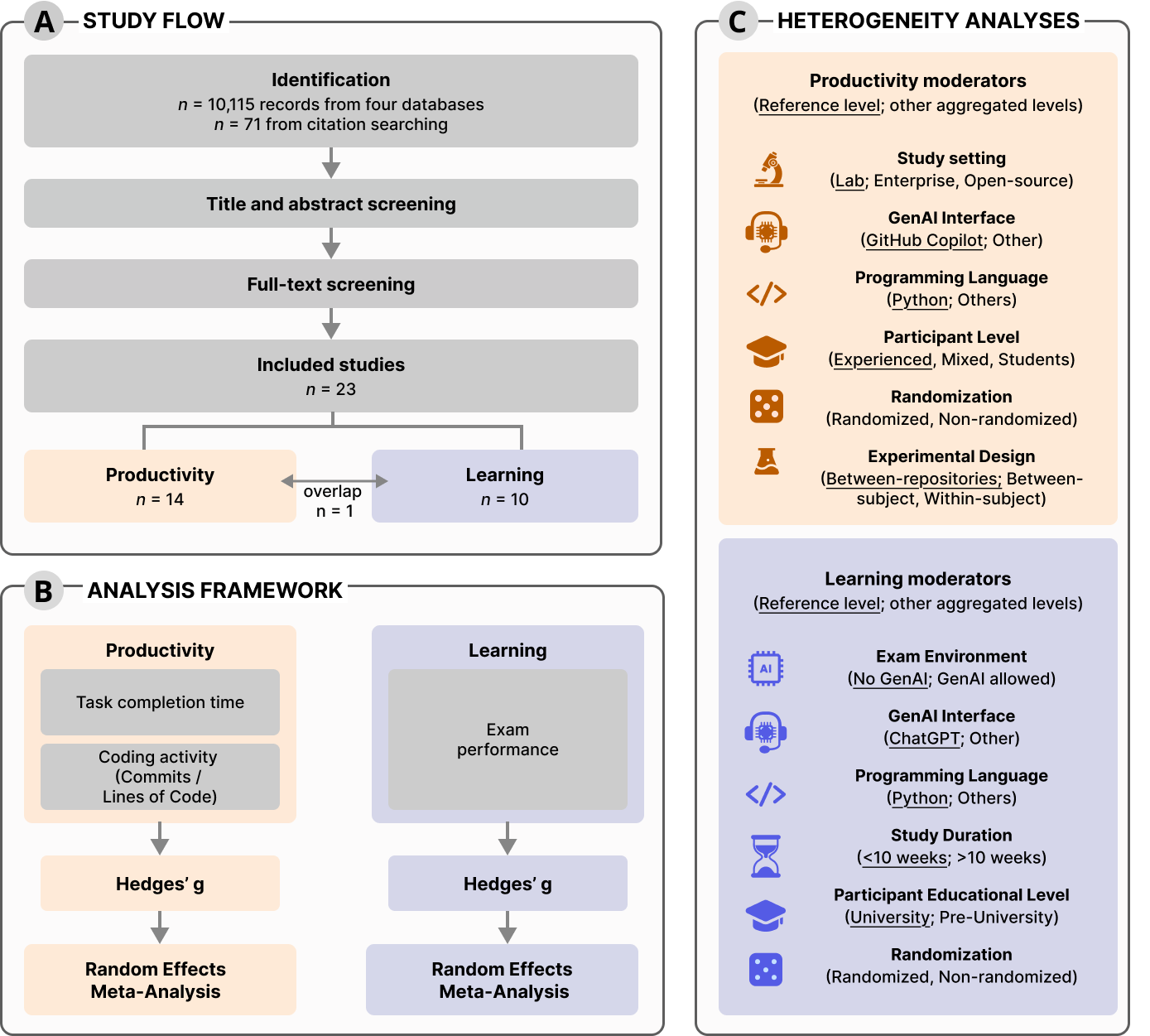}
\caption{\textbf{Research overview.} \textbf{a,}~Overview of the systematic process for literature review and data synthesis. \textbf{b,}~The meta-analysis estimates the effect of GenAI on productivity (measured by task completion time and code output) and learning (measured by exam performance) across the identified studies. \textbf{c,}~Moderator analyses examine how study context, task type, and assessment conditions contribute to variation across reported effect sizes. This allows to identify when and where GenAI assistance is effective.}
\label{fig:overview}
\end{figure}

Here, we conduct a pre-registered meta-analysis to quantify the impact of GenAI on productivity and learning (see Figure~\ref{fig:overview}). After screening $10{,}115$ records during our literature search, we identified $n = 23$ relevant studies. We extracted quantitative measures of productivity and learning from each study and then computed standardized effect sizes based on Hedges’ $g$ under a random-effects model. To explain variability in outcomes across studies, we conducted moderator analyses to examine differences across study context, participant characteristics, and technological implementation. This approach allows us to identify when and where GenAI assistance is effective.

\newpage
\section*{Results}
\label{sec:results}

\subsection*{Productivity effect of GenAI (RQ1)}

\begin{figure}[H]
    \centering
    \includegraphics[
        width=1\textwidth,
        trim=0cm 0cm 0cm 0.8cm, 
        clip
    ]{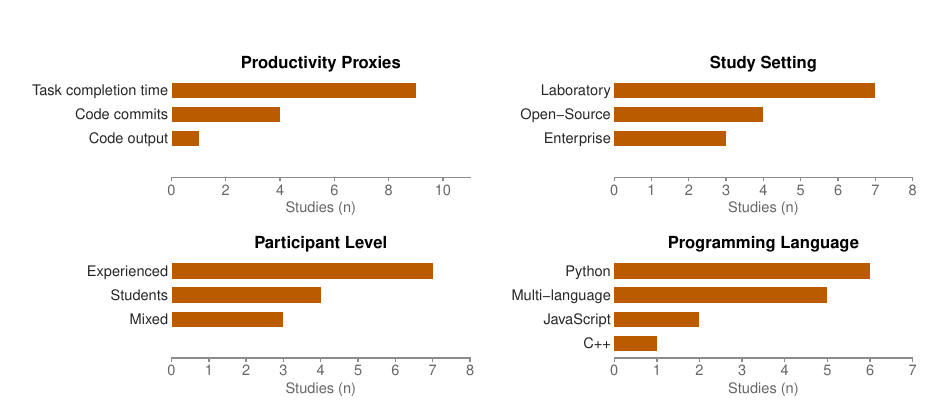}
    \caption{\textbf{Characteristics of productivity studies.}}
    \label{fig:productivity_study_overview}
\end{figure}

\subsubsection*{Study overview}

To analyze the effect of GenAI on productivity (RQ1), we conducted a systematic synthesis of the existing literature (see Method section). Overall, our search identified 16 effect size estimates derived from $n = 14$ unique studies that examine the impact of GenAI assistance on developer productivity (see~\Cref{fig:overview} for a descriptive overview and \Cref{tab:productivity_studies} for the full list of studies). The productivity studies comprise studies on a participants level ($m = 3535$ participants in total) as well as studies conducted on a repository level ($r = 6355$ repositories in total). The studies operationalized productivity using common proxies for measuring development activities in software engineering research (see~\Cref{fig:productivity_study_overview}), including task completion time ($n = 9$), the number of commits ($n = 4$), and the lines of output code ($n = 1$). The studies span open-source ($n = 4$), enterprise ($n = 3$), and controlled laboratory settings ($n = 7$). The studies were published between 2022 and 2025, and include both peer-reviewed publications ($n = 6$) and preprints ($n = 8$), which reflect the rapid pace of GenAI research. 

The research methods of the studies vary and include randomized controlled trials ($n = 10$), natural experiments ($n = 2$), and quasi-experiments ($n = 2$). Evidence is generated based on within-subject comparisons ($n = 6$), between-repository comparisons ($n = 3$), and between-subject comparisons ($n = 5$).

The primary GenAI assistant was GitHub Copilot ($n = 6$ studies), while the other studies used tools such as Cursor ($n = 2$) and other GenAI-based systems ($n = 6$). Python was the most frequently used language ($n = 6$), with the remaining studies conducted in multi-language settings ($n = 5$), JavaScript ($n = 2$), or C++ ($n = 1$). Taken together, the existing studies focusing on productivity effects reveal substantial heterogeneity in the study design. 

\subsubsection*{Main meta-analysis}

To examine how GenAI assistance affects programmer productivity (RQ1), we synthesized evidence from existing studies using a random-effects meta-analysis. Specifically, we consolidated the 16 independent effect size estimates that compared GenAI-assisted development with a baseline condition involving manual programming into a single pooled estimate, while accounting for variation across study contexts.

\begin{figure}[H]
    \centering
    \includegraphics[
        width=0.7\textwidth,
        trim=0cm 0cm 0cm 1cm, 
        clip
    ]{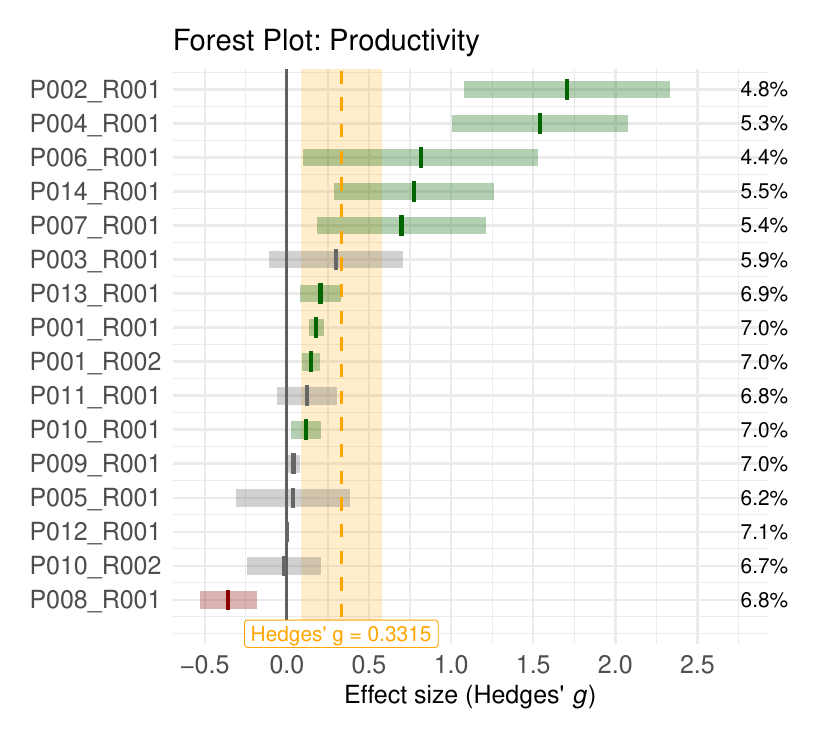}
    \caption{\textbf{Forest plot of the pooled effect of GenAI assistance on developer productivity.} To answer RQ1, the plot summarizes individual effect size estimates (Hedges' $g$) and the corresponding $95\%$ CIs, with study weights shown on the right. The vertical line at $g = 0$ denotes a null effect, while estimates to the right indicate higher productivity gain from GenAI assistance. The orange dashed line shows the pooled estimate, and the orange shaded area represents the $95\%$ CI of the pooled estimate. Overall, the meta-analysis indicates a moderate positive effect (Hedges’ $g = 0.33$, $95\%$ CI: $[0.09, 0.58]$, $\mathrm{SE} = 0.13$, $p = 0.008$), alongside substantial heterogeneity across studies ($I^2 = 99\%$).}
    \label{fig:prod_forest_plot}
\end{figure}

Overall, we find a statistically significant, positive but moderate increase in productivity associated with GenAI assistance (Hedges’ $g = 0.33$, $95\%$ confidence interval [CI]: $[0.09, 0.58]$, $\mathrm{SE} = 0.13$, $p = 0.008$). The forest plot is shown in \Cref{fig:prod_forest_plot}. At the same time, productivity effects vary substantially, which is reflected by a substantial heterogeneity across studies ($I^2 = 99\%$), with an estimated between-study variance of $\tau^2 = 0.22$ ($Q(15) = 206.06$, $p < 0.001$). This suggests that GenAI assistance does not yield uniform productivity gains, but that effects depend strongly on where and how these tools are used.

\subsubsection*{Heterogeneity in productivity effects}

\begin{figure}[H]
    \centering
    \includegraphics[
        width=1\textwidth,
        trim=0cm 0cm 0cm 0.8cm, 
        clip
    ]{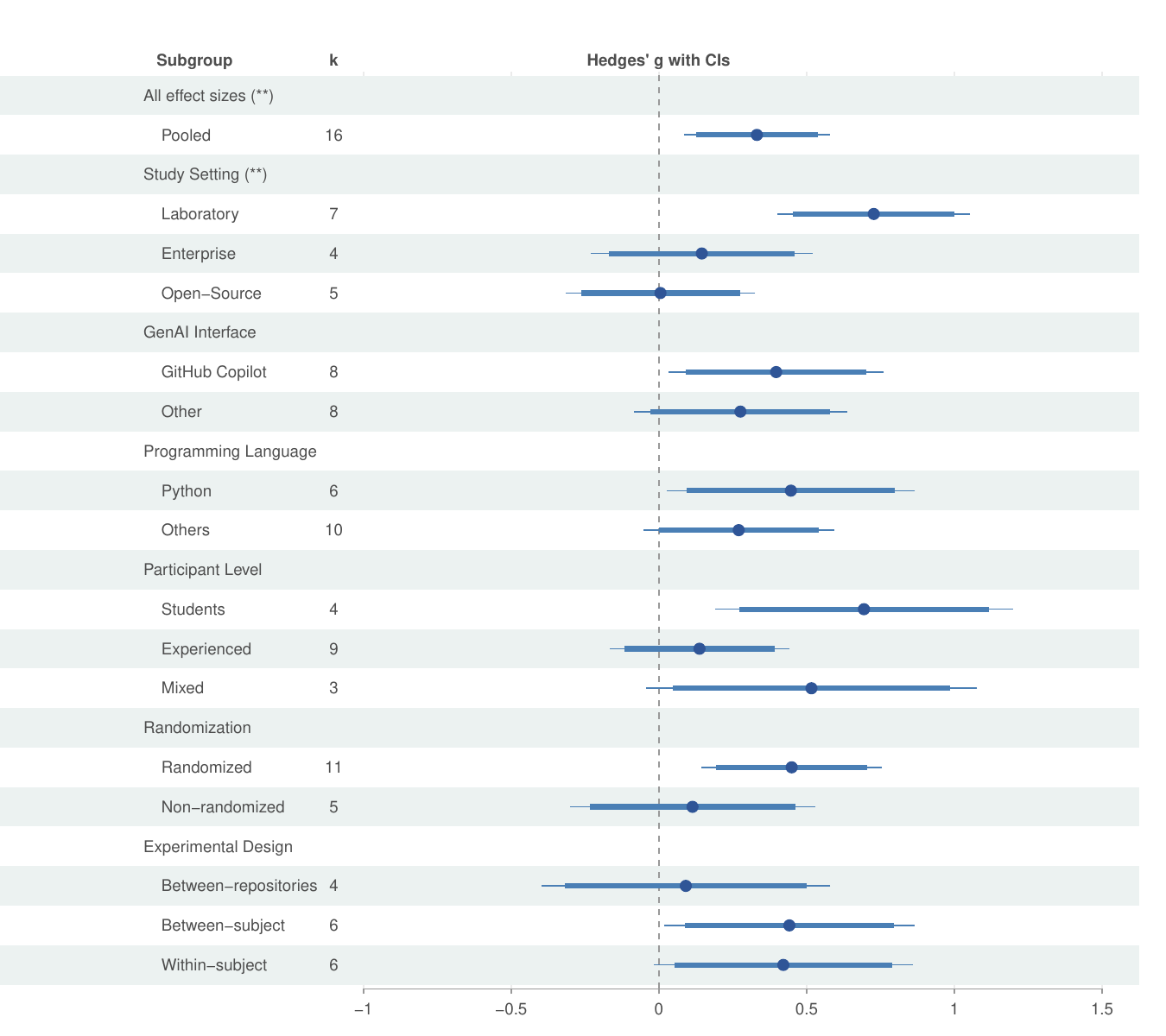}
   \caption{\textbf{Predicted productivity effect sizes by moderator subgroup.} 
Predicted effect sizes (Hedges' $g$) derived from univariate random-effects 
meta-regressions (REML estimator). Thick bars represent 90\% confidence 
intervals; thin bars represent 95\% confidence intervals. $k$ denotes the 
number of effect sizes per subgroup. Asterisks in parentheses after moderator 
category labels indicate statistically significant omnibus tests 
($^*p < 0.05$; $^{**}p < 0.01$; $^{***}p < 0.001$). The vertical dashed line 
represents a null effect ($g = 0$); positive values indicate a beneficial 
effect of GenAI assistance on productivity.}
    \label{fig:mod_analyses_prod}
\end{figure}

To understand when and where GenAI is effective, we conducted moderator analyses to identify key determinants of productivity gains. Specifically, we analyzed six moderators: (a)~study setting (i.e., laboratory, enterprise, open-source), (b)~the GenAI interface (e.g., GitHub Copilot), (c)~programming language (e.g., Python), (d)~the participant level (students, experienced, mixed), (e)~whether the study was randomized (randomized, non-randomized), and (f)~the experimental design (between-repositories, between-subject, within-subject). The moderators were selected to reflect commonly reported study characteristics, while ensuring sufficient representation and thus statistical power across subgroups. All moderator analyses were performed using mixed-effects meta-regression models; the results are summarized in \Cref{fig:mod_analyses_prod}.

We examined whether the context in which GenAI tools are evaluated influences observed productivity gains by comparing controlled laboratory experiments, enterprise settings, and open-source environments. The study setting may shape outcomes because tasks in controlled laboratory experiments are typically shorter, more structured, and require less familiarization with existing code bases, whereas real-world development involves coordination overhead, constraints imposed by existing and legacy code bases, and quality assurance. Differences may further arise between enterprise and open-source settings due to different collaboration practices, governance structures, and programmer backgrounds. Interestingly, the moderation analysis reveals that the study setting significantly moderates the productivity effect of GenAI ($Q_M(2) = 10.51$, $p = 0.005$), accounting for approximately 36\% of the heterogeneity across studies. Laboratory experiments yield a large and significant effect ($g = 0.73$, $p < 0.001$), whereas effects in enterprise ($g = 0.19$, $p = 0.448$) and open-source settings ($g = 0.01$, $p = 0.975$) are substantially smaller and do not differ significantly from zero. This result suggests that the productivity gains with GenAI tools observed in experiments diminish substantially in more realistic contexts.

None of the remaining moderators reached statistical significance at common significance thresholds, which is the case for the GenAI interface ($Q_M(1) = 0.21$, $p = 0.644$), programming language ($Q_M(1) = 0.43$, $p = 0.514$), and the participant level ($Q_M(2) = 3.97$, $p = 0.138$). Also, in terms of study design, neither randomization ($Q_M(1) = 1.63$, $p = 0.202$), nor experimental design was a significant moderator ($Q_M(2) = 1.35$, $p = 0.510$), suggesting that the substantial residual heterogeneity observed across studies could not be accounted for by whether studies were randomized or not and employed a  between-repositories, within-subject, or between-subject design. However, these results should be interpreted cautiously, given that the small number of studies and limited variability across moderator levels may have been insufficient to reliably detect meaningful differential effects.

\subsubsection*{Time trend of productivity effects}

\begin{figure}[H]
    \centering
    \includegraphics[
        width=.7\textwidth,
        trim=0cm 0cm 3cm 1.1cm, 
        clip
    ]{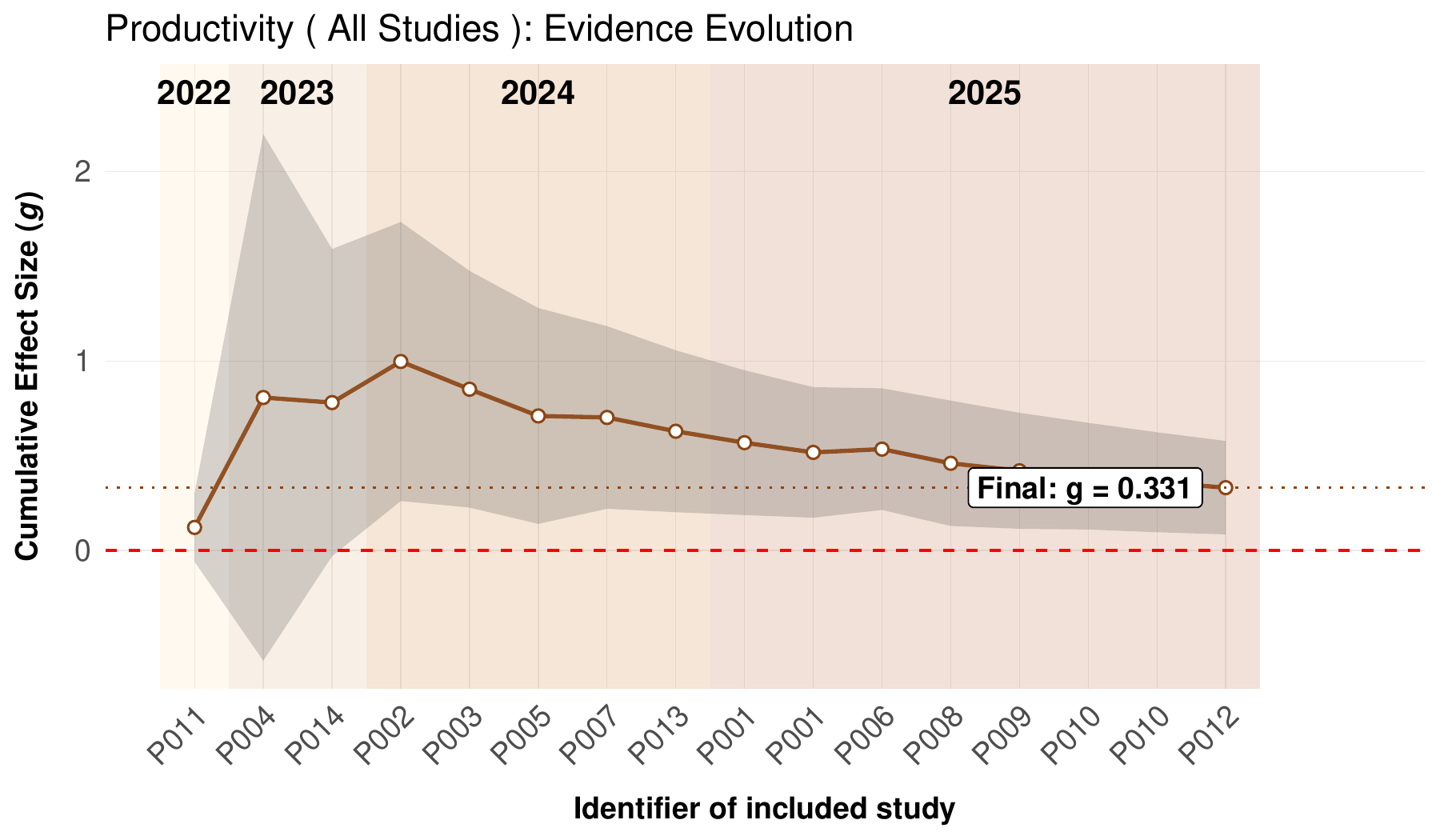}
    \caption{\textbf{Longitudinal analysis of productivity effects by publication year.} The plot shows a cumulative meta-analysis over time. Each point shows the pooled effect size after sequentially adding studies in chronological order from 2022 to 2025, while background colors indicate the publication year. Shaded bands represent $95\%$ CIs around the cumulative estimates. Light background shading distinguishes different time periods. The horizontal dashed line (in red) denotes a null effect, and the dotted line represents the final pooled estimate of the overall effect. Study identifiers on the x-axis (e.g., P012, P005) correspond to the order in which studies and the respective effect sizes enter the cumulative analysis.}
    \label{fig:productivity_evolution_plot}
\end{figure}

To evaluate how the reported productivity effects of GenAI have changed over time, we conducted a cumulative meta-analysis ordered by publication year (see \Cref{fig:productivity_evolution_plot}). In the cumulative analysis, studies are chronologically ordered and then added one by one to the meta-analysis to update the pooled estimate with the overall effect size from all studies up to a specific year. This approach thus tracks how the estimated productivity effect evolves as new studies are published. 

The cumulative meta-analysis shows a consistent, modest productivity gain that remains stable as studies are added. Despite the rapid advancement of underlying foundation models over this period, the cumulative evidence does not reflect a corresponding increase in reported productivity gains. Rather, the cumulative estimate stabilized at a small, consistent productivity gain ($g = 0.331$) as evidence accumulated. This suggests that raw model capability or AI literacy may not be the primary drivers of productivity gains in GenAI-assisted coding tasks.

\subsubsection*{Robustness checks}

We conducted several further analyses as robustness checks. We examined whether effect sizes differed between studies rated as lower versus higher risk of bias using RoB2 and ROBINS-I (see Online Supplement~\ref{appendix:RoB}). Risk of bias was not a significant moderator ($Q_M(1) = 2.90$, $p = 0.089$). However, lower-risk studies showed descriptively smaller effect sizes than higher-risk studies ($g = 0.08$, 95\% CI: $[-0.29, 0.45]$ vs.\ $g = 0.50$, 95\% CI: $[0.19, 0.80]$).

The leave-one-out sensitivity analysis showed that the pooled effect size was robust to the exclusion of individual studies (range: $g = 0.24$--$0.38$, all $p < 0.05$), but where the omission of P002 \cite{weber_significant_2024} and P004 \cite{kazemitabaar_studying_2023} produced the largest reductions. Influence diagnostics further identified these two studies as exerting disproportionate influence on the pooled estimate, evidenced by Cook's $d$ values ($0.55$ and $0.56$) and studentized residuals exceeding the conventional $|z| > 3$ threshold ($3.01$ and $2.83$), suggesting their effect sizes deviated substantially from the remaining studies. However, as the effect remained positive and statistically significant across all leave-one-out iterations, these studies did not alter the overall interpretation. Notably, heterogeneity remained consistently high across all iterations ($I^2 = 98.8\text{--}99.6\%$), thereby confirming that no single study accounted for the substantial between-study variance.

We further examined the potential presence of publication bias and small-study effects using Egger’s mixed-effects regression test \cite{egger_bias_1997}, which indicated significant funnel plot asymmetry ($z = 4.52, p < 0.001$). To account for this asymmetry, we applied the trim-and-fill procedure \cite{duval_trim_2000}, which imputed no missing studies, suggesting that the detected asymmetry does not reflect a systematic suppression of null results. This could imply that the funnel plot asymmetry is due to true heterogeneity rather than publication bias per se.

\subsection*{Learning effect from GenAI (RQ2)}

\subsubsection*{Study overview}

\begin{figure}[H]
    \centering
    \includegraphics[
        width=1\textwidth,
        trim=0cm 0cm 0cm 0.8cm, 
        clip
    ]{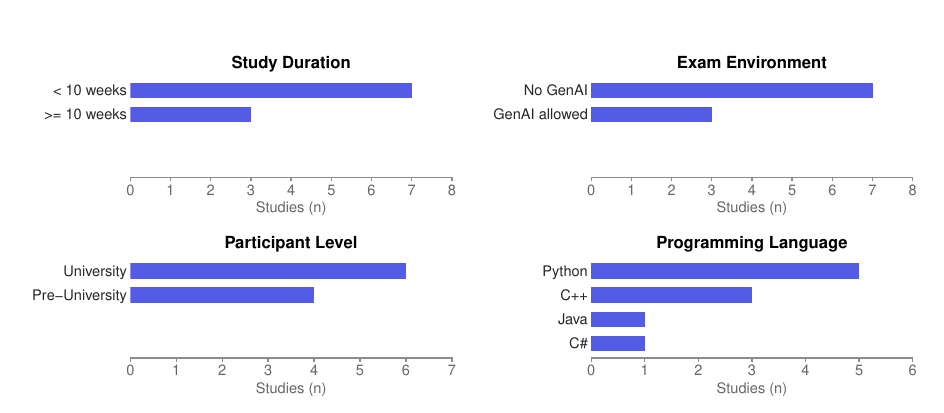}
    \caption{\textbf{Characteristics of learning effect studies.}}
    \label{fig:learn_study_overview}
\end{figure}

To analyze the effect of GenAI on learning (RQ2), we again conducted a systematic literature search to then perform a meta-analysis (see Method section). Overall, we identified $n = 10$ unique studies examining the impact of GenAI on programming-related learning outcomes, comprising $m = 1,069$ participants in total, from which we extracted 11 effect size estimates (see \Cref{fig:learn_study_overview} for a descriptive overview and \Cref{tab:learning_studies} for the full list of studies). All effect sizes are based on quantitative measures of learning outcomes, specifically comparisons of exam performance following instruction phases in which students either had access to GenAI assistants or worked without them. The studies were published between 2023 and 2025, all in peer-reviewed outlets, and span a geographically diverse set of countries, including Slovenia, South Korea, China, Taiwan, Oman, and the United States. 

Methodologically, $n = 4$ studies employed experimental designs, while $n = 6$ studies used quasi-experimental designs to compare exam performance between subjects. Most studies are based on samples of university students ($n = 6$), while the remaining studies involve high school ($n = 2$), middle school ($n = 1$), and mixed K-12 samples ($n = 1$).

Variation in GenAI tools was narrow, with OpenAI models accounting for the majority of deployments across all but two studies. Python was the most commonly used programming language ($n = 5$), followed by C++ ($n = 3$), Java ($n = 1$), and C\# ($n = 1$). The study duration also varied substantially: $n = 7$ studies were from GenAI interventions lasting fewer than ten weeks, while $n = 3$ studies were based on studies where the duration of the GenAI intervention exceeded ten weeks. In most cases, GenAI assistance was not allowed during final assessments ($n = 7$), whereas $n = 3$ studies did allow the usage of GenAI in the treatment group for the assessment.

\subsubsection*{Main meta-analysis}

To examine how GenAI assistance affects learning outcomes in programming education (RQ2), we synthesized evidence from 11 independent effect sizes from $n = 10$ studies that compared GenAI-supported learning with traditional educational approaches. For this, we again use a random-effects model while accounting for variation across education contexts. Notably, the included studies differed in whether GenAI assistance was available during the assessment itself or only during the learning phase, a distinction that, as we show in the section below, proves critical for interpreting the pooled results.

\begin{figure}[H]
    \centering
    \includegraphics[
        width=0.7\textwidth,
        trim=0cm 0cm 0cm 1cm,
        clip
    ]{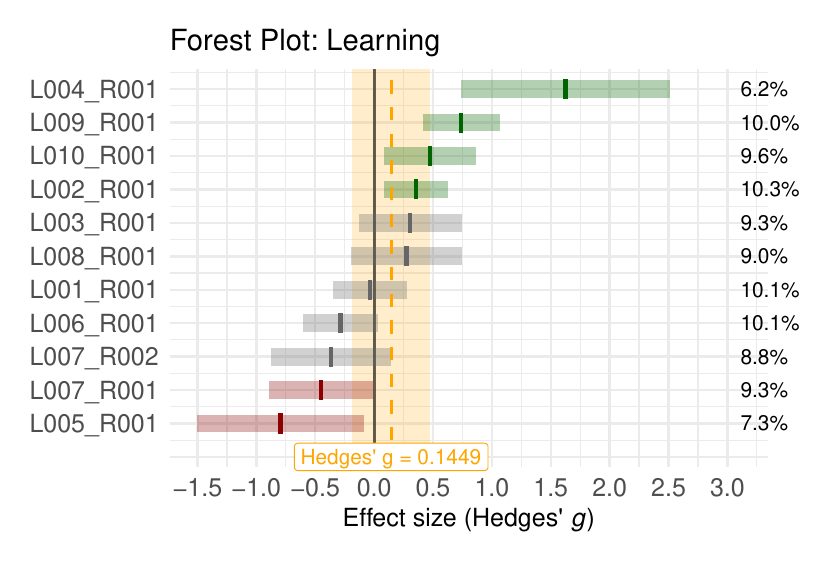}
    \caption{\textbf{Forest plot of the pooled effect of GenAI assistance on learning outcomes (RQ2).} Rows represent the individual effect size estimates from different studies, together with the $95\%$ CIs as shaded areas and study weights shown on the right. The vertical line at $g = 0$ denotes a null effect, while estimates to the right indicate a higher learning effect from GenAI assistance. The orange dashed line shows the pooled estimate (Hedges' $g$) from random-effects meta-analysis, and the orange shaded area represents the $95\%$ CI of the pooled estimate. Overall, the pooled effect size is small and statistically non-significant (Hedges’ $g = 0.14$, $\mathrm{SE} = 0.17$, $p = 0.389$, $95\%$ CI $[-0.18, 0.47]$), alongside substantial heterogeneity across studies ($I^2 = 86\%$).}
    \label{fig:learn_forest_plot}
\end{figure}

Overall, we find a small but statistically non-significant pooled effect of GenAI assistance on learning outcomes (Hedges’ $g = 0.14$, $95\%$ CI: $[-0.18, 0.47]$, $\mathrm{SE} = 0.17$, $p = 0.389$). The corresponding CI includes zero, indicating no reliable overall improvement in exam performance associated with GenAI use across the analyzed studies. The forest plot is shown in \Cref{fig:learn_forest_plot}. At the same time, substantial heterogeneity was observed, with an $I^2$ statistic of $86\%$ and an estimated between-study variance of $\tau^2 = 0.25$ ($Q(10) = 54.96$, $p < 0.001$). This level of heterogeneity suggests that the effects of GenAI assistance on learning outcomes depend strongly on the contextual setting, as for example on whether students retained access to GenAI during the test.


\subsubsection*{Heterogeneity in learning effects}
\label{learning_heterogeneity}

\begin{figure}[H]
    \centering
    \includegraphics[
        width=1\textwidth,
    ]{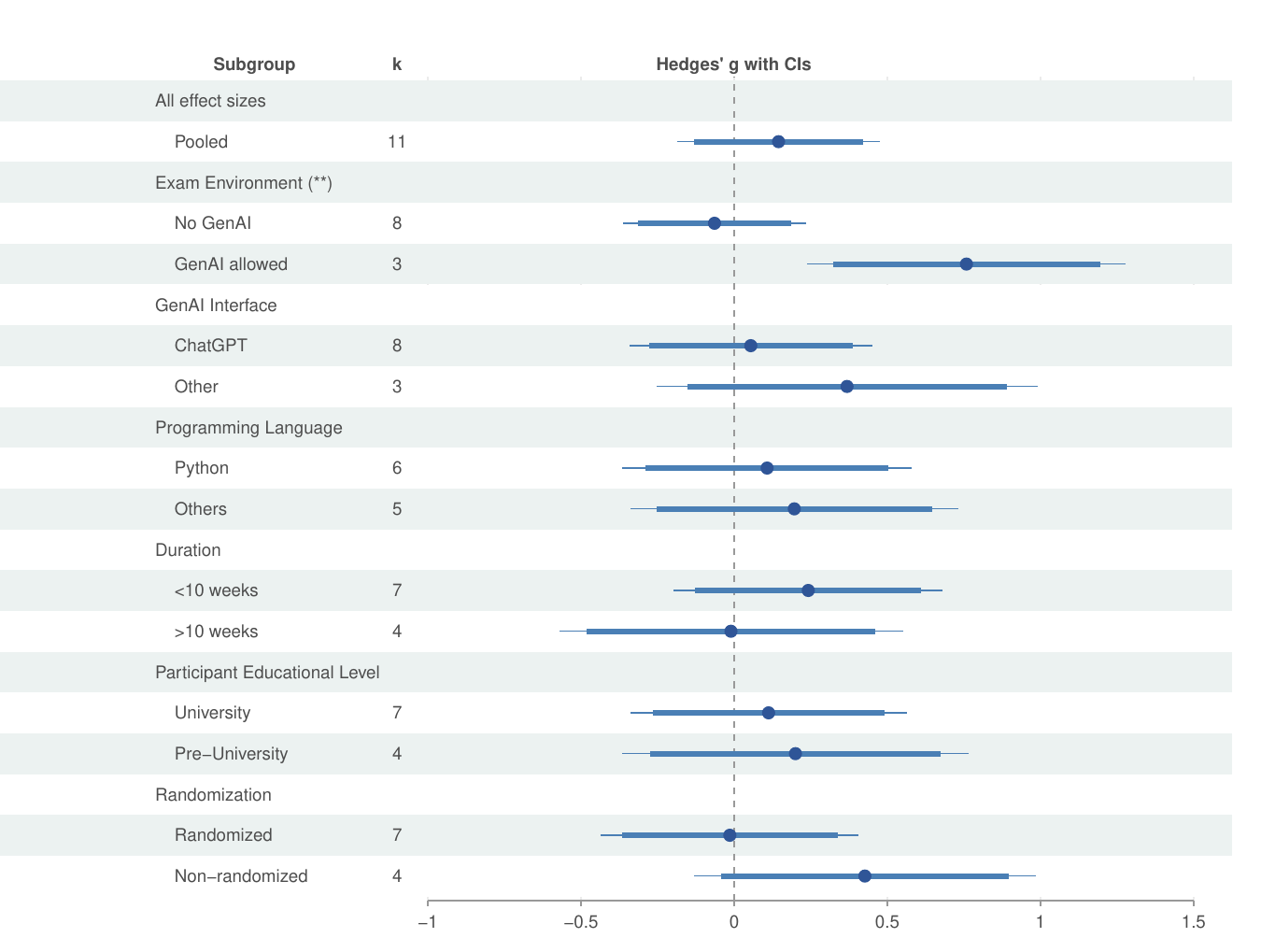}
    \caption{\textbf{Predicted learning effect sizes by moderator subgroup.} Predicted effect sizes (Hedges' $g$) derived from univariate random-effects meta-regressions (REML estimator). Thick bars represent 90\% confidence intervals; thin bars represent 95\% confidence intervals. $k$ denotes the number of effect sizes per subgroup. Asterisks in parentheses after moderator category labels indicate statistically significant omnibus tests ($^*p < 0.05$; $^{**}p < 0.01$; $^{***}p < 0.001$). The vertical dashed line represents a null effect ($g = 0$); positive values indicate a beneficial effect of GenAI assistance on learning outcomes.}
    \label{fig:learning_study_overview}
\end{figure}

To understand when and where GenAI affects learning outcomes in programming education, we conducted univariate moderator analyses to identify key sources of variation across studies. Again, the moderator analyses were performed using mixed-effects meta-regression models; the results are summarized in \Cref{fig:learning_study_overview}. 

The most influential moderator was the exam environment, meaning whether GenAI assistance was permitted during assessment. The overall test of moderation was significant ($Q_M(1) = 7.23$, $p = 0.007$), indicating that the testing environment explained between-study heterogeneity. Indeed, the moderator accounted for $45.49\%$ of the total heterogeneity, representing a substantial reduction from the unconditional model. When participants were not allowed to use AI for the assessment, the effect was non-significant and even pointed slightly in a negative direction ($g$ = $-0.06$, 95\% CI: [$-0.36$, $0.24$], $p = 0.674$). In contrast, when AI tools were permitted during the assessment, the estimated effect was large and statistically significant ($g$ = $0.76$, $95\%$ CI: [$0.24$, $1.28$], $p = 0.004$). This pattern suggests that students perform better on assessments when AI tools are available, but these performance gains may reflect tool-dependent augmentation rather than durable learning, raising concerns about the transferability of GenAI-assisted instruction to unaided contexts.

None of the remaining moderators significantly explained variation in learning effect sizes. The type of GenAI interface did not moderate the effect ($Q_M(1) = 0.70$, $p = 0.404$), nor did the programming language taught ($Q_M(1) = 0.06$, $p = 0.807$). Similarly, study duration ($Q_M(1) = 0.48$, $p = 0.487$), and the participant educational level (university vs.\ pre-university; $Q_M(1) = 0.06$, $p = 0.812$) were all non-significant moderators. Further, as for the productivity outcomes, the randomization was not a significant moderator ( $Q_M(1) = 1.53$, $p = 0.216$). Notably, none of these five moderators accounted for any of the observed heterogeneity ($R^2 = 0.00\%$ in each case), and residual heterogeneity remained substantial throughout ($I^2 > 85\%$). This suggests that, among the study-level characteristics we coded, only the exam environment accounts for a meaningful share of the observed heterogeneity.

\subsubsection*{Robustness checks}

To assess robustness, we first performed a leave-one-out sensitivity analysis, which indicates that the pooled effect size was generally robust to the exclusion of individual studies, with estimates ranging from $g = 0.06$ to $g = 0.21$. Importantly, no single study altered the overall conclusion. Additionally, and consistent with the productivity analysis, risk of bias was not a significant moderator of learning effects ($Q_M(1) = 1.15$, $p = 0.284$; see Online Supplement~\ref{appendix:robustness_checks}). Further, to assess potential publication bias, we conducted Egger's regression test for funnel plot asymmetry \cite{egger_bias_1997} and a trim-and-fill analysis \cite{duval_trim_2000}. Egger's test revealed no significant asymmetry ($z = 0.55$, $p = 0.585$), and the trim-and-fill procedure estimated zero missing studies, leaving the pooled estimate unchanged. Together, these results provide no indication of publication bias in the learning outcome. Similarly, influence diagnostics suggest that the learning dataset is relatively robust to outliers as no study exceeded conventional thresholds.

\newpage
\section*{Discussion}
\label{sec:discussion}


\subsection*{Summary of findings}

Our meta-analysis demonstrates that GenAI-based coding assistants can improve developer productivity, yet the evidence does not support a corresponding benefit for learning outcomes in programming education. This finding challenges overly optimistic claims from artificial benchmark results suggesting universal productivity gains \cite{chen2021evaluating, jimenez2023swe, jain2024livecodebench}, while the broader meta-analytic evidence also counters the notion that current GenAI tools are ineffective in practice \cite{becker_measuring_2025}. Instead, the strong variation across studies indicates that the impact of GenAI assistance depends heavily on contextual characteristics. For productivity (RQ1), GenAI assistance is associated with a moderate positive effect, though gains are substantially larger in controlled laboratory environments than in open-source or real-world settings, suggesting that the simplified tasks in experimental settings may amplify measured effects relative to everyday development workflows. For learning (RQ2), significant effects emerged only when students retained access to GenAI during the assessment itself, suggesting that observed gains reflect tool-augmented performance rather than durable skill acquisition. This finding aligns with recent evidence that the educational benefits of GenAI only materialize when it is integrated deliberately into the learning process, as unrestricted access may hinder genuine skill acquisition \cite{bastaniGenerativeAIGuardrails2025c, shenHowAIImpacts2026a}.

\subsection*{Implications}


Our analysis provides robust evidence that GenAI-based coding assistants can improve developer productivity, but the benefits depend strongly on context. Productivity gains are larger in controlled laboratory settings than in open-source or real-world development environments. This suggests that results from controlled experiments may overestimate the productivity improvements that organizations can expect in everyday software development, where code bases are more complex and coordination costs are higher \cite{cui_effects_2024, shihab_effects_2025, Xu_aiassisteddecrease2025, paradis_how_2024}. For practice, this means that GenAI tools should not be treated as a one-size-fits-all productivity solution. Instead, the specific design choices of how GenAI tools are integrated into existing workflows, review processes, and quality controls \cite{simkute_ironies_2024, banh_copiloting_2025, russo_manifesto_2024} may matter just as much as the technical capabilities of the models alone. As GenAI tools take over more routine code generation \cite{jin_can_2024}, developers may spend less time writing code and more time reviewing, validating, and integrating AI-generated outputs \cite{bird_taking_2023, Xu_aiassisteddecrease2025, becker_measuring_2025}. Companies should therefore be careful when generalizing results from experimental studies to production environments \cite{paradis_how_2024}.
In this context, emerging agentic workflows that better align GenAI assistance with developer roles and responsibilities may be particularly beneficial. Overall, our results suggest that the long-term value of GenAI in software engineering will depend not only on continued advances in model capability but also considerably on how effectively organizations embed these tools within real-world development processes. This points to a needed shift in focus away from simple coding benchmarks and toward a more human-centered understanding of how these tools can complement developer expertise and intent \cite{collinsBuildingMachinesThat2024b, wangposition}.


While GenAI tools are now widely used among computer science students, our analysis does not support a robust improvement in learning outcomes. Although students perform better on assessments when allowed to use AI, these benefits do not transfer to settings without AI support. This suggests that GenAI may help students complete tasks without internalizing the underlying reasoning. When students rely on GenAI to generate or repair code, they may bypass critical learning processes such as reasoning through errors and internalizing core concepts. This pattern mirrors previous research in other educational domains \cite{piscitelli_influence_2024, bastaniGenerativeAIGuardrails2025c, becker_programming_2023} and is consistent with concerns about cognitive offloading \cite{prather_robots_2023, vivian_cognitiveoffload_2025, jose_cognitive_2025}; as a result, such reliance on GenAI tools may backfire when students are required to explain, adapt, or debug AI-generated code. 

These risks are particularly salient given ongoing shifts in software engineering practice, where developers increasingly take on roles that emphasize code review, integration, and oversight rather than routine code production \cite{acharya_generative_2025, lee_impact_2025}. If students graduate without strong foundations in code comprehension and reasoning, they may struggle in these higher-level roles. At the same time, banning GenAI tools is neither realistic nor desirable given the growing role of GenAI in professional programming practice \cite{bardach_bridging_2025, beale_computer_2025}. Paradoxically, the very skills that GenAI makes more important in professional practice, critical evaluation, code reasoning, and independent problem solving, are precisely those that may atrophy when students rely on GenAI during learning \cite{shenHowAIImpacts2026a}. This tension is amplified by the fact that novice developers, although they might experience immediate productivity gains from GenAI-assisted prototyping and debugging \cite{peng_impact_2023, cui_effects_2024}, often lack the understanding needed to critically evaluate generated code \cite{barkeGroundedCopilotHow2023, guHowAnalystsUnderstand2024}. Instead, experienced developers might leverage these tools differently by using them to offload routine implementation while redirecting their effort toward complex problem solving \cite{hoffmann2025generative, daniottiWhoUsingAI2026}. Our findings thus suggest the need to rethink curricula and the integration of GenAI by moving beyond code production alone and placing greater emphasis on explanation, reasoning, and critical evaluation, so that students develop the skills required to work safely and effectively with GenAI assistance \cite{kirova_software_2024,  rahe_how_2025, ma_scaffolding_2025, denny_computing_2024}.

\subsection*{Current state of the literature}

Our meta-analysis provides a thorough analysis on the effect of GenAI assistance on productivity and learning outcomes by deliberately focusing on comparable outcomes across various settings. However, as with other studies, ours is subject to limitations that reflect the early and rapidly evolving nature of research on GenAI-assisted coding. First, GenAI capabilities evolved rapidly during the observation period, yet our cumulative meta-analysis shows that reported productivity effects remained stable over time rather than increasing with new model improvements. 

Second, with 16 effect sizes for productivity and 11 for learning, the evidence base constrains the granularity of moderator analyses. This reflects both the emerging state of the field and our deliberate decision to exclude studies relying on subjective productivity measures, which capture a fundamentally different construct \cite{meyer_software_2014}. We therefore reduced the number of covariates and relied on univariate meta-regressions, where the risk of overfitting would outweigh interpretive gains.

Third, as preregistered, we included both randomized and non-randomized studies, which relies on the assumption that both are comparable \cite{borenstein_conversions2009}. Further, the actual GenAI use was often not controlled for, which implies that the study design captures treatment assignment, not adherence to GenAI use. Additionally, factors such as repository size, coordination overhead, legacy code, quality assurance processes, and the specific mode of interaction with GenAI might be relevant characteristics for productivity and learning outcomes, yet remain largely unobserved in the available evidence.

\subsection*{Recommendations for future research}

Our literature synthesis also highlights impactful opportunities for future research. For productivity (RQ1), more emphasis should be placed on field studies conducted under realistic development conditions. In contrast to controlled laboratory settings, professional software development involves large code bases with legacy code, coordination overhead, and strict quality and maintainability requirements. Studying GenAI-assisted coding in such environments is important to better assess the external validity of productivity effects and to explain the gap between laboratory and real-world findings observed in our meta-analysis. Such analyses should explicitly account for key confounding factors, including team structure, project complexity, and organizational processes. Second, future research should investigate the interaction formats and how to implement GenAI tools in systematic comparisons across different GenAI technologies, including reasoning-capable GenAI, agentic systems, and emerging practices such as ``vibe coding'', to better understand how GenAI capabilities and usage patterns influence productivity. Third, future research would benefit from a deeper understanding of developer heterogeneity. Separating expertise (e.g., novices, junior and senior roles) from task type (e.g., code review, quality assurance, and maintenance) would help resolve this, while also clarifying whether GenAI primarily accelerates routine work or enables developers to reallocate attention to higher-order activities such as architectural design and code review \cite{gambacorta2024generative}.

Similarly, for learning outcomes (RQ2), future research should focus on how GenAI tools can be leveraged as effective tutoring systems for programming skill acquisition. In particular, studies should investigate the mechanisms and boundary conditions through which GenAI assistance influences learning. Experimental designs that systematically vary how GenAI outputs are presented, such as providing full solutions versus guardrails that present structured hints or constructive feedback \cite{bastaniGenerativeAIGuardrails2025c}, could help distinguish when GenAI encourages superficial problem solving and when it supports conceptual understanding. Moreover, research should examine how GenAI assistance can be adaptively designed to meet the needs of users across different expertise levels, from novice programmers to experienced developers, accounting for individual differences in learning styles and user needs. Finally, longitudinal studies are needed to assess the long-term effects of GenAI assistance on learning, knowledge retention, and skill development, including potential risks from deskilling after the adoption of GenAI by experienced developers.

GenAI is driving a major transformation in software development by shifting work from manually writing code toward reviewing and integrating AI-generated output. As these tools become increasingly embedded in software development workflows, a key tension emerges: productivity gains do not necessarily translate into stronger human skill development and may, in some cases, undermine it. Addressing this tension is important for computer science education, given that effective human oversight will become a key skill in GenAI-enabled programming.


\newpage
\section*{Methods}
\label{sec:methods}

\subsection*{Search strategy}

Our data collection process follows the PRISMA 2020 guidelines \cite{Page_Prisma2020} for systematic literature reviews. The meta-analysis was preregistered on December~4, 2025 using the PRISMA-P protocol (see \url{https://osf.io/5h2s3}).

We operationalized productivity and learning as follows to ensure consistency across the meta-analysis: (1)~Productivity was defined as the efficiency of code production \cite{ziegler_productivity_2022, peng_impact_2023, cui_effects_2024}. In the included studies, this was measured via task completion time in a controlled laboratory experiment, the number of commits, or the number of merged pull requests. (2)~Learning was defined as the measurable acquisition of programming skills. In the context of software engineering education, this is primarily measured via exam scores following a learning phase with or without GenAI support \cite{kazemitabaar_studying_2023, choi_impact_2025}.

We performed a systematic search across four major electronic databases commonly used to disseminate research findings in computer science and software engineering: (i)~ACM Digital Library, (ii)~arXiv, (iii)~Scopus, and (iv)~Web of Science. To account for the rapid evolution of generative AI research, our search included both peer-reviewed publications and preprints to capture emerging research that may not yet be formally published in peer-reviewed conferences or journals. In addition, we performed backward and forward citation searches for all included studies to identify additional relevant publications not retrieved by the initial search query. The search was restricted to publications in English language released between January~1, 2019, and December~5, 2025, which includes the public release of GenAI-based coding assistants such as GitHub Copilot but also research based on earlier large language models such as GPT-2. 

We formulated two search queries to identify studies examining the effects of GenAI assistance on (1) developer productivity and (2) learning outcomes. Both queries were structured around three building blocks that filtered studies along a specific dimension: (i)~the technology (i.e., GenAI, large language models, \ldots), (ii)~the application context (i.e., programming, software engineering, \ldots), and (iii)~the outcome of interest, namely, either productivity or learning. Within each building block, we used a set of synonyms, including common abbreviations, to ensure that the search is sufficiently broad to cover relevant studies. The formulation was inspired by prior research aimed at analyzing the effects of GenAI on human behavior in other contexts \cite{holblingMetaanalysisPersuasivePower2025, holznerGenerativeAICreativity2025} but was tailored to programming. The boolean search strings are shown below, while the exact versions adjusted to the specific databases are provided in Online~Supplement~\ref{appendix:search_queries}.

\begin{tcolorbox}[
    colframe=darkheader,         
    colback=lightbody,          
    coltitle=white,             
    title=\textbf{Search query (Productivity)},
    listing only,
    fonttitle=\bfseries\small,  
    fontupper=\ttfamily\footnotesize\singlespacing,  
    arc=1.5mm,                    
    boxrule=0.5mm,
    top=-6mm,
    bottom=0mm 
]
\begin{lstlisting}[
  breaklines=true,
  breakatwhitespace=true,
  xleftmargin=0em,
  columns=fullflexible]
TITLE("GitHub Copilot" OR "Copilot" OR "CodeWhisperer" OR "Gemini" OR "Claude" OR "ChatGPT" OR "Large Language Model" OR "LLM" OR "AI" OR "Generative AI" OR "GenAI" OR "AI programming assistant" OR "programming tool" OR "AI code assistant") 
AND TITLE-ABS-KEY("developer" OR "software engineer" OR "programmer" OR "software development") 
AND TITLE-ABS-KEY("productivity" OR "efficiency" OR "experience" OR "behavior" OR "human-AI collaboration")
\end{lstlisting}
\end{tcolorbox}

\begin{tcolorbox}[
    colframe=darkheader,         
    colback=lightbody,          
    coltitle=white,             
    title=\textbf{Search query (Learning)},
    listing only,
    fonttitle=\bfseries\small,  
    fontupper=\ttfamily\footnotesize\singlespacing, 
    arc=1.5mm,                    
    boxrule=0.5mm,  
    top=-6mm,
    bottom=0mm 
]
\begin{lstlisting}[
  breaklines=true,
  breakatwhitespace=true,
  xleftmargin=0em,
  columns=fullflexible]
TITLE-ABS-KEY("GitHub Copilot" OR "Copilot" OR "CodeWhisperer" OR "Gemini" OR "Claude" OR "GPT" OR "ChatGPT" OR "Large Language Model" OR "LLM" OR "LLMs" OR "programming assistant" OR "code assistant" OR "code completion" OR "AI-assisted" OR "Chatbot")
AND (("student*" OR "novice*" OR "learner*" OR "junior*") AND ("developer*" OR "programmer*"))
\end{lstlisting}
\end{tcolorbox}


Eligible studies had to meet the following inclusion criteria: (i)~the study compared human performance with GenAI assistance against a control condition without GenAI assistance or a valid alternative baseline; (ii)~the study reported empirical, quantitative results; (iii)~the study measured at least one relevant outcome variable related to developer productivity (e.g., task completion time, output volume) or learning (e.g., test or exam scores); and (iv)~the study provided sufficient statistical information to compute standardized effect sizes (e.g., means, standard deviations, sample sizes), or such information could be reliably gathered from supplementary materials. Screening and eligibility assessment were conducted by the first author. 

\begin{landscape}
\begin{figure}[htbp]
    \centering
    \makebox[\textwidth][c]{
        \includegraphics[
            width=1.28\textwidth,
            height=\textheight,
            keepaspectratio,        
            trim=1.2cm 3cm 1.2cm 1.8cm, 
            clip
        ]{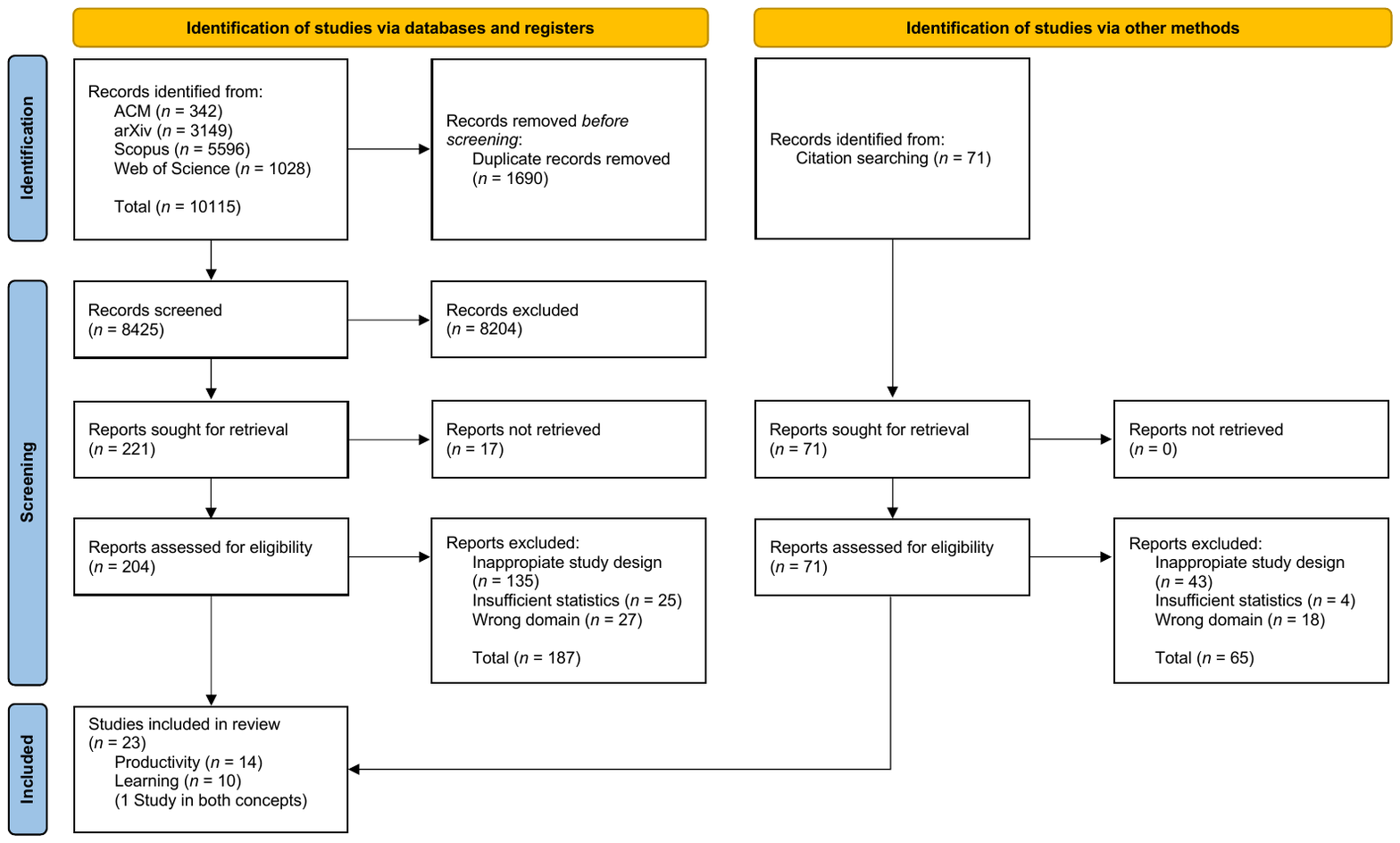}
    }
    \caption{\textbf{Study inclusion flowchart.} The diagram shows the study selection process following the PRISMA 2020 guidelines \cite{Page_Prisma2020}.}
    \label{fig:prisma_flow}
\end{figure}
\end{landscape}

    
The inclusion flowchart is shown in \Cref{fig:prisma_flow}. Our database search yielded a total of $n = 10{,}115$ records from ACM ($n = 342$), arXiv ($n = 3{,}149$), Scopus ($n = 5{,}596$), and Web of Science ($n = 1{,}028$). We identified duplicate records based on (i) the DOI and (ii) a combination of the first author's surname and the title. For the latter, we used a normalized format where all special characters and whitespace were removed prior to comparison. After removing $n = 1{,}690$ duplicate records, $n = 8{,}425$ records remained for title and abstract screening. Of these, $n = 8{,}204$ records were excluded due to a lack of relevance to the topic of our meta-analysis. The full texts of the remaining $n = 221$ reports were sought for retrieval, of which $n = 17$ could not be accessed, leaving $n = 204$ reports for eligibility assessment. During this stage, $n = 135$ reports were excluded due to insufficient study design, $n = 25$ due to insufficient statistical reporting, and $n = 27$ because the study addressed a different domain, totaling to $n = 187$ ineligible reports. When studies were otherwise eligible but lacked sufficient statistical information, we contacted the corresponding authors via email to request the necessary data. None provided the information needed for inclusion. Consequently, $n = 17$ studies retrieved from the four databases met our inclusion criteria.

To minimize the risk of missing relevant works, we additionally conducted citation-based searches (forward and backward) for the studies retrieved from the databases, which identified $n = 71$ further records. All retrieved studies were assessed for eligibility, resulting in the exclusion of $n = 65$ studies, either due to inappropriate study design ($n = 43$), insufficient statistics ($n = 4$), or an incorrect domain ($n = 18$). 

In total, $n = 23$ studies met all inclusion criteria and were retained for the meta-analysis. Of these, $n = 13$ studies examined only productivity outcomes, $n = 9$ focused solely on learning outcomes, and $n = 1$ study reported on both types of outcomes.

\subsection*{Data collection}

All studies that met the predefined inclusion criteria ($n = 23$) were manually transferred into a spreadsheet. Data extraction was performed by the first and the second author. The lists of included studies for productivity and learning are provided in \Cref{tab:productivity_studies} and \Cref{tab:learning_studies}, respectively. The complete dataset with the extracted statistics is available in our GitHub repository as supplementary material (see the data and code availability statement). For each study, we extracted metadata (e.g., author, title, publication year, publication stage), details about the experimental design (e.g., sample size, study design, choice of GenAI assistant, programming language), and outcome statistics (e.g., mean, standard deviation, $t$-value, $F$-value, standardized $\beta$). The variables were extracted to explore heterogeneity across different contexts (see Online Supplement~\ref{appendix:definitions}).

\subsection*{Statistical analysis}

To compare the statistical results across the included studies, we computed effect sizes using Hedges' $g$. We aimed to extract Cohen’s $d$ directly from the studies where possible, yet most studies only reported other statistical measures (means, standard deviations, sample sizes, $t$-values, $F$-values), based on which we computed Cohen’s $d$ using established conversion formulas (see Online Supplement~\ref{appendix:es_conversions}). For studies employing a pre-post-control group design, we computed the effect size $d_{\mathrm{ppc2}}$ as recommended by Morris~\cite{morrisEstimatingEffectSizes2008}, defined as the difference in mean pre-post change between the treatment and control groups, divided by the pooled pretest standard deviation. To integrate effect sizes from studies using within-subject designs with those using between-subject designs, we followed the framework of Morris and DeShon~\cite{morrisCombiningEffectSize2002}. Where the pre-post correlation was not reported, we assumed $r = 0.5$, a conservative default consistent with recommendations in the literature~\cite{morrisCombiningEffectSize2002}. For the detailed conversion formulae, see Online Supplement~\ref{appendix:es_conversions}.
The bias-correction factor $J$ was applied to all estimates to mitigate small-sample bias \cite{hedges_distribution_1981}. In the main paper, all point estimates are reported with their corresponding $95\%$ confidence interval (CI) and standard error (SE).

When studies reported multiple objective outcome measures within the same outcome category, we selected a single representative effect size per study to preserve statistical independence. For productivity outcomes, we prioritized commit-based measures, as these reflect discrete, externally-verifiable metrics that measure the contribution to software development. Measures based on lines of code were omitted whenever alternatives were available, as such measures may be inflated by redundant code generated by GenAI (e.g., comments or other form of in-code documentation) and tend to be less directly tied to meaningful development output. For learning outcomes, most studies reported test-based performance measures. When both intermediate assessments (e.g., midterms) and final exams were available, we selected the final exam score, as these represent a more comprehensive assessment of learning over the full instructional period.

In line with Cochrane's recommendations \cite{higgins_cochrane_2024}, we employed a random-effects model, as the included studies vary considerably in their design, context, and methodology, making the assumption of a single common effect implausible. We estimated the pooled effect sizes using the Restricted Maximum Likelihood (REML) estimator \cite{Viechtbauer_REML2005} and $95\%$ CIs, which represent the expected range of true effects. Separate random-effects models were fitted for productivity-related outcomes and learning-related outcomes. Between-study heterogeneity was assessed using Cochran’s $Q$ test, the $I^2$ statistic to estimate the proportion of total variance attributable to heterogeneity, and the estimated between-study variance ($\tau^2$) \cite{borenstein_conversions2009}.

We conducted subgroup analyses and meta-regressions to explore systematic sources of heterogeneity from moderators such as the study setting, the programming language, etc. The moderators differed between productivity and learning, reflecting the specific context and available data in each domain.

All data processing and statistical modeling were performed in R (version 4.5.1) using the \texttt{metafor} package (version 4.8.0) \cite{Viechtbauer_metaregression2010}. The data and code to replicate our analyses are available via our GitHub repository (see the data and code availability statement).

\subsection*{Robustness checks}

We performed several robustness checks following best practice. First, we performed leave-one-out sensitivity analyses, in which the meta-analysis was re-estimated repeatedly while sequentially excluding one study at a time. This procedure evaluates whether the overall results are driven by any single study and helps identify influential studies that disproportionately affect the pooled effect size or heterogeneity. In addition, we computed influence diagnostics following established procedures \cite{Viechtbauer_influencediag2010}.

Potential publication bias was evaluated using funnel plot asymmetry and Egger’s regression test \cite{egger_bias_1997}, conditional on a minimum of ten effect sizes per analysis, following Cochrane's guidance \cite{higgins_cochrane_2024}. We applied the trim-and-fill procedure  \cite{duval_trim_2000} to estimate the number and impact of potentially missing studies due to selective reporting. To address the statistical non-independence introduced by multiple effect sizes drawn from the same study \cite{cheung2014modeling}, we additionally estimated a three-level meta-analytic model with effect sizes nested within studies (see~\Cref{tab:rob_multilevel} in Online Supplement~\ref{appendix:robustness_checks}).

\subsection*{Study quality}
We evaluated study quality by assessing the risk of bias of each effect size individually. Given the diverse range of study designs included in this meta-analysis, a single risk-of-bias tool would not have adequately captured the methodological variability across studies. We therefore selected the most appropriate tool for each study design: ROBINS-I for non-randomized studies ($n = 9$), and the appropriate RoB-2 variant for randomized studies, choosing between the individually randomized parallel group trials ($n = 11$), crossover ($n = 5$), and cluster ($n = 1$) versions depending on the specific design (see Tables~\ref{tab:productivity_studies} and~\ref{tab:learning_studies}). This ensured that the risk-of-bias evaluation was appropriately adapted to the methodological characteristics of each study design.

In our case, two raters independently assessed the risk of bias of each study by answering the corresponding questions and building an algorithmic overall evaluation. Differences were then discussed and a consensus reached. We used the risk of bias ratings to evaluate whether preprints, that are not yet peer-reviewed, account for a higher portion in high risk studies. Second, we categorized studies as either lower risk (low or some concerns) or higher risk (high or critical) and conducted a moderator analysis to evaluate whether risk-of-bias level was associated with differences in effect size estimates. 

Across both productivity and learning outcomes, the overall study quality was mixed (see Online Supplement~\ref{appendix:RoB}). Of all effect sizes, 59.3\% were rated as having a higher risk of bias, whereas 40.7\% had a lower risk of bias. Common issues were deviations from intended interventions for randomized studies, within-subject design studies that did not account for carry-over effects, and the widespread absence of preregistration. We conducted a univariate moderation analysis to check whether the overall risk of bias moderates the effect (see Online Supplement~\ref{appendix:robustness_checks}). Risk of bias was not a statistically significant moderator for either productivity ($Q_M = 2.90$, $p = 0.089$) or learning ($Q_M = 1.15$, $p = 0.284$), indicating that the findings are robust regardless of study quality. Descriptively, however, higher risk studies tended to report larger effect sizes than lower risk studies for both outcomes.

\begin{landscape}
    \begin{table}[ht]
        \centering
        \scriptsize
        \singlespacing 
        
        \rowcolors{2}{gray!10}{white}
        
        \begin{tabularx}{\linewidth}{l|l|X|l|l|l|l|l|l|l|l|l}
            
            \rowcolor{gray!30}
            \textbf{ID} & \textbf{RID} & \textbf{Study} & \textbf{Lang.} & \textbf{GenAI Asst.} & \textbf{Base LLM} & \textbf{Setting} & \textbf{Outcome} & \textbf{Unit} & \textbf{Peer-Reviewed} & \textbf{Bias Tool} & \textbf{Overall Bias} \\
            \hline
            P001 & R001 & Yeverechyahu et al.\ \cite{yeverechyahu_impact_2024} & Python & Copilot & GPT-3 & Open-source & Commits & Count/Qtr & No & ROBINS-I & Moderate \\
            
            P001 & R002 & Yeverechyahu et al.\ \cite{yeverechyahu_impact_2024} & Rust & Copilot & GPT-3 & Open-source & Commits & Count/Qtr & No & ROBINS-I & Moderate \\
            
            P002 & R001 & Weber et al.\ \cite{weber_significant_2024} & Python & Copilot & GPT-3 & Laboratory & Time & Seconds & Yes & RoB 2 (co) & High \\
            
            P003 & R001 & Paradis et al.\ \cite{paradis_how_2024} & C++ & -- & -- & Enterprise & Time & Minutes & No & RoB 2 (IRPG) & Some concerns \\
            
            P004 & R001 & Kazemitabaar et al.\ \cite{kazemitabaar_studying_2023} & Python & Codex & Davinci & Laboratory & Time & Seconds & Yes & RoB 2 (IRPG) & High \\
            
            P005 & R001 & Nam et al.\ \cite{nam_using_2024} & Python & GILT & 3.5-turbo & Laboratory & Time & Sec/Task & Yes & RoB 2 (co) & High \\
            
            P006 & R001 & Shihab et al.\ \cite{shihab_effects_2025} & Mixed & Copilot & -- & Laboratory & Time & Seconds & Yes & RoB 2 (co) & High \\
            
            P007 & R001 & Piscitelli et al.\ \cite{piscitelli_influence_2024} & JS & ClueBot & GPT-4o & Laboratory & Time & Seconds & Yes & RoB 2 (co) & High \\
            
            P008 & R001 & Becker et al.\ \cite{becker_measuring_2025} & Python & Cursor & Sonnet & Open-source & Time & \% Change & No & RoB 2 (IRPG) & Some concerns \\
            
            P009 & R001 & Xu et al.\ \cite{Xu_aiassisteddecrease2025} & Mixed & Copilot & GPT-3 & Open-source & Commits & Log Count & No & ROBINS-I & Moderate \\
            
            P010 & R001 & Cui et al.\ \cite{cui_effects_2024} & Mixed & Copilot & GPT-3 & Enterprise & Commits & Count/Wk & No & RoB 2 (IRPG) & High \\
            
            P010 & R002 & Cui et al.\ \cite{cui_effects_2024} & Mixed & Copilot & GPT-3 & Enterprise & Commits & Count/Wk & No & RoB 2 (IRPG) & High \\
            
            P011 & R001 & Xu et al.\ \cite{xu_-ide_2022} & Python & NL2Code & -- & Laboratory & Time & Seconds & Yes & RoB 2 (co) & High \\
            
            P012 & R001 & He et al.\ \cite{he_does_2025} & Mixed & Cursor & -- & Open-source & Commits & Log Count & No & ROBINS-I & Serious \\

            P013 & R001 & Gambacorta et al.\ \cite{gambacorta2024generative} & Mixed & CodeFuse & -- & Enterprise & Code Output & Lines of Code & No & ROBINS-I & Moderate \\

            P014 & R001 & Peng et al.\ \cite{peng_impact_2023} & JavaScript & Copilot & Codex & Laboratory & Time & Minutes & No & RoB 2 (IRPG) & High \\
            
            \hline
            
        \end{tabularx}
        \caption{\textbf{Overview of included studies related to productivity.} \\ 
        \textit{Note:} \textbf{ID} = Unique identifier for each study; \textbf{RID} = Unique identifier for each effect size estimate derived from the study. \\
        \textbf{Bias Tool}: RoB 2 = Revised Cochrane Risk-of-Bias tool (IRPG = individually randomized parallel-group; co = crossover trial); ROBINS-I = Risk Of Bias In Non-randomised Studies -- Interventions. \textbf{Overall Bias} = Overall risk-of-bias judgment.}
        \label{tab:productivity_studies}
    \end{table}
    \end{landscape}

\begin{landscape}
    \begin{table}[ht]
        \centering
        \scriptsize
        \singlespacing 
        
        \rowcolors{2}{gray!10}{white}
        
        \begin{tabularx}{\linewidth}{l|l|X|l|l|l|l|l|l|l|l}
            
            \rowcolor{gray!30}
            \textbf{ID} & \textbf{RID} & \textbf{Study} & \textbf{Lang.} & \textbf{Setting} & \textbf{Dur.} & \textbf{Exam} & \textbf{Outcome} & \textbf{Unit} & \textbf{Bias Tool} & \textbf{Overall Bias} \\
            \hline
            L001 & R001 & Kosar et al.\ \cite{kosar_computer_2024} & C++ & Course & >10w & No GenAI & Exam performance & 0-100 & RoB 2 (IRPG) & High \\
            
            L002 & R001 & Choi \& Kim \cite{choi_impact_2025} & Python & Course & <10w & No GenAI & Exam performance & 0-100 & ROBINS-I & Moderate \\
            
            L003 & R001 & Sun et al.\ \cite{sun_would_2024} & Python & Course & <10w & GenAI allowed & Exam performance & 0-100 & RoB 2 (IRPG) & High \\
            
            L004 & R001 & Abdulla et al.\ \cite{abdulla_using_2024} & C\# & Course & <10w & GenAI allowed & Exam performance & 0-100 & ROBINS-I & Critical \\
            
            L005 & R001 & Johnson et al.\ \cite{johnson_using_2024} & C++ & Laboratory & <10w & No GenAI & Exam performance & 0-100 & RoB 2 (IRPG) & High \\
            
            L006 & R001 & Yang et al.\ \cite{yang_effectiveness_2025} & C++ & Course & <10w & No GenAI & Exam performance & 0-100 & ROBINS-I & Moderate \\
            
            L007 & R001 & Suh et al.\ \cite{suh_programming_2025} & Python & Course & >10w & No GenAI & Exam performance & 0-10 & RoB 2 (IRPG) & Some concerns \\
            
            L007 & R002 & Suh et al.\ \cite{suh_programming_2025} & Python & Course & >10w & No GenAI & Exam performance & 0-10 & RoB 2 (IRPG) & Some concerns \\
            
            L008 & R001 & Kazemitabaar et al.\ \cite{kazemitabaar_studying_2023} & Python & Laboratory & <10w & No GenAI & Exam performance & 0-100 & RoB 2 (IRPG) & High \\
            
            L009 & R001 & Fan et al.\ \cite{fan_impact_2025} & Java & Course & >10w & GenAI allowed & Exam performance & 0-100 & RoB 2 (cluster) & High \\
            
            L010 & R001 & Tang et al.\ \cite{tangEnhancingProgrammingPerformance2025} & Python & Course & <10w & No GenAI & Exam performance & n.s. & ROBINS-I & Moderate \\
            
            \hline
       
        \end{tabularx}
        \caption{\textbf{Overview of included studies related to learning.} \\
        \textit{Note:} \textbf{ID} = Unique identifier for each study; \textbf{RID} = Unique identifier for each effect size estimate derived from the study. \\
        \textbf{Setting}: Course = Course-integrated. \\
        \textbf{Bias Tool}: RoB 2 = Revised Cochrane Risk-of-Bias tool (IRPG = individually randomized parallel-group; cluster = cluster-randomized); ROBINS-I = Risk Of Bias In Non-randomised Studies -- Interventions. \textbf{Overall Bias} = Overall risk-of-bias judgment.}
        \label{tab:learning_studies}
    \end{table}
    \end{landscape}

\newpage

\vspace{0.4cm}
\section*{Author contributions} 
S.M. and M.G. contributed equally to this work. S.M. and S.F. had the idea. M.G. conducted the initial literature search. S.M. and M.G. performed study screening, risk-of-bias assessment, and statistical analyses. All authors contributed to conceptualization, manuscript writing, and approved the final manuscript.

\vspace{0.4cm}
\section*{Funding}

S.F. acknowledges funding via the Swiss National Science Foundation (SNSF), Grants 197485 and 186932.

\vspace{0.4cm}
\section*{Competing interests}
The authors declare no competing interests.

\vspace{0.4cm}
\phantomsection
\section*{Data and code availability} 

The data and the code to replicate our analyses are available via our Git repository at \url{https://github.com/SM2982/MetaanalysisGenAICoding.git}. The repository also includes the PRISMA checklist.


\newpage
\bibliography{literature}


\newpage

\appendix

\section{Search Queries}
\label{appendix:search_queries}

{
\singlespacing  
\scriptsize   
\setlength\tabcolsep{4pt} 


\begin{longtable}{@{} p{0.08\textwidth} p{0.10\textwidth} >{\raggedright\arraybackslash}p{0.76\textwidth} @{}}

    \caption{\textbf{Search strings.} Below are the search strings adapted for the different databases.} \label{tab:search_strings} \\
    \toprule
    \textbf{Database} & \textbf{Topic} & \textbf{Query} \\
    \midrule
    \endfirsthead
    
    \caption[]{(continued)} \\
    \toprule
    \textbf{Database} & \textbf{Topic} & \textbf{Query} \\
    \midrule
    \endhead
    
    \bottomrule
    \endfoot


   
    \textbf{Scopus}
    & Productivity 
    & TITLE("GitHub Copilot" OR "Copilot" OR "CodeWhisperer" OR "Gemini" OR "Claude" OR "ChatGPT" OR "Large Language Model" OR "LLM" OR "AI" OR "Generative AI" OR "GenAI" OR "AI programming assistant" OR "programming tool" OR "AI code assistant") 
    AND TITLE-ABS-KEY("developer" OR "software engineer" OR "programmer" OR "software development") 
    AND TITLE-ABS-KEY("productivity" OR "efficiency" OR "experience" OR "behavior" OR "human-AI collaboration") \\ 
    \addlinespace 
    
    & Learning 
    & TITLE-ABS-KEY("GitHub Copilot" OR "Copilot" OR "CodeWhisperer" OR "Gemini" OR "Claude" OR "GPT" OR "ChatGPT" OR "Large Language Model" OR "LLM" OR "LLMs" OR "programming assistant" OR "code assistant" OR "code completion" OR "AI-assisted" OR "Chatbot") 
    AND (("student*" OR "novice*" OR "learner*" OR "junior*") AND ("developer*" OR "programmer*")) \\
    \midrule

    \textbf{Web of Science}
    & Productivity 
    & TI=("GitHub Copilot" OR "Copilot" OR "CodeWhisperer" OR "Gemini" OR "Claude" OR "ChatGPT" OR "Large Language Model" OR "LLM" OR "AI" OR "Generative AI" OR "GenAI" OR "AI programming assistant" OR "programming tool" OR "AI code assistant")
    AND TS=("developer" OR "software engineer" OR "programmer" OR "software development")
    AND TS=("productivity" OR "efficiency" OR "experience" OR "behavior" OR "human-AI collaboration") \\
    \addlinespace
    
    & Learning 
    & TS=(("GitHub Copilot" OR "Copilot" OR "CodeWhisperer" OR "Gemini" OR "Claude" OR "GPT" OR "ChatGPT" OR "Large Language Model" OR "LLM" OR "LLMs" OR "programming assistant" OR "code assistant" OR "code completion" OR "AI-assisted" OR "Chatbot") AND (("student*" OR "novice*" OR "learner*" OR "junior*") AND ("developer*" OR "programmer*"))) \\
    \midrule

    \textbf{ACM}
    & Productivity 
    & "query": \{ Title:("GitHub Copilot" OR "Copilot" OR "CodeWhisperer" OR "Gemini" OR "Claude" OR "ChatGPT" OR "Large Language Model" OR "LLM" OR "AI" OR "Generative AI" OR "GenAI" OR "AI programming assistant" OR "programming tool" OR "AI code assistant" OR "AI-powered programming") AND (Title:("developer" OR "software engineer" OR "programmer" OR "software development") OR Abstract:("developer" OR "software engineer" OR "programmer" OR "software development")) AND (Title:("productivity" OR "efficiency" OR "experience" OR "behavior" OR "human-AI collaboration") OR Abstract:("productivity" OR "efficiency" OR "experience" OR "behavior" OR "human-AI collaboration")) \} \newline "filter": \{ \} \\
    \addlinespace
    
    & Learning 
    & "query": \{ Title:("GitHub Copilot" OR "Copilot" OR "CodeWhisperer" OR "Gemini" OR "Claude" OR "ChatGPT" OR "Large Language Model" OR "LLM" OR "AI" OR "Generative AI" OR "GenAI" OR "AI programming assistant" OR "programming tool" OR "AI code assistant" OR "AI-powered programming") AND (Abstract:("developer" OR "software engineer" OR "programmer" OR "software development") OR Title:("developer" OR "software engineer" OR "programmer" OR "software development")) AND (Abstract:("productivity" OR "efficiency" OR "experience" OR "behavior" OR "human-AI collaboration") OR Title:("productivity" OR "efficiency" OR "experience" OR "behavior" OR "human-AI collaboration")) \} \newline "filter": \{ \} \\
    \midrule

    \textbf{arXiv}
    & Productivity 
    & Title:(\$Github Copilot\$) OR Title: (Copilot OR Gemini OR Claude OR GPT OR Cursor OR LLM OR AI)
    AND Abstract: (developer OR programmer OR software development)
    AND Abstract: (productivity OR efficiency) \\
    \addlinespace
    
    & Learning 
    & Title:(\$Github Copilot\$) OR Title: (Copilot OR Gemini OR Claude OR GPT OR ChatGPT OR AI OR LLM OR Chatbot)
    AND Abstract: (student OR novice OR learner OR junior)
    AND Abstract: (developer OR programmer) \\

\end{longtable}
}


\newpage

\section{Robustness Checks}
\label{appendix:robustness_checks}
\begin{table}[ht]
\centering
\caption{\textbf{Moderator analysis results for risk of bias across productivity 
and learning outcomes.} Predicted effect sizes (Hedges' $g$) with 95\% confidence 
intervals derived from univariate mixed-effects meta-regressions (REML estimator). 
$k$ = number of effect sizes per subgroup. $Q_M$ and $p$ reflect the omnibus 
moderator test.}
\label{tab:rob_moderator}
\resizebox{\textwidth}{!}{%
\begin{tabular}{lcccccccccc}
\toprule
& \multicolumn{5}{c}{\textbf{Productivity}} 
& \multicolumn{5}{c}{\textbf{Learning}} \\
\cmidrule(lr){2-6} \cmidrule(lr){7-11}
\textbf{Subgroup} & $k$ & $g$ & \textbf{95\% CI} & $Q_M$ & $p$ 
                  & $k$ & $g$ & \textbf{95\% CI} & $Q_M$ & $p$ \\
\midrule
\textbf{Risk of Bias} & & & & & & & & & & \\
\quad Lower risk  
  & 6 & 0.081 & {[}$-$0.287,\ 0.449{]} & \multirow{2}{*}{2.90} & \multirow{2}{*}{0.089} 
  & 5 & $-$0.043 & {[}$-$0.521,\ 0.434{]} & \multirow{2}{*}{1.15} & \multirow{2}{*}{0.284} \\
\quad Higher risk 
  & 10 & 0.498 & {[}0.190,\ 0.805{]} & & 
  & 6 & 0.319 & {[}$-$0.140,\ 0.778{]} & & \\
\bottomrule
\end{tabular}%
}
\end{table}

\begin{table}[ht]
\centering
\caption{\textbf{Multilevel meta-analysis accounting for effect size dependencies.} Results from three-level meta-analytic models with effect sizes (level 1) nested within studies (level 2), estimated via REML. $\sigma^2_1$ = between-study variance; $\sigma^2_2$ = within-study variance. Within-study variance is estimated at zero for both outcomes, confirming that the pooled estimates and conclusions from the main random-effects analysis remain robust when accounting for dependencies among effect sizes from the same study.}
\label{tab:rob_multilevel}
\resizebox{\textwidth}{!}{%
\begin{tabular}{lcccccccc}
\toprule
\textbf{Outcome} & $k$ & $g$ & \textbf{SE} & \textbf{95\% CI} & $p$ & $\sigma^2_1$ & $\sigma^2_2$ & $Q$ \\
\midrule
Productivity & 16 & 0.38 & 0.15 & {[}0.09,\ 0.67{]} & 0.010 & 0.266 & 0.000 & 206.06*** \\
Learning     & 11 & 0.19 & 0.18 & {[}$-$0.154,\ 0.54{]} & 0.275 & 0.261 & 0.000 & 54.96*** \\
\bottomrule
\multicolumn{9}{l}{\footnotesize $^{***}p < .001$; $k$ = number of effect sizes.} \\
\end{tabular}%
}
\end{table}

\newpage
\section{Risk of Bias}
\label{appendix:RoB}

\begin{figure}[htbp]
    \centering
    \includegraphics[width=\textwidth]{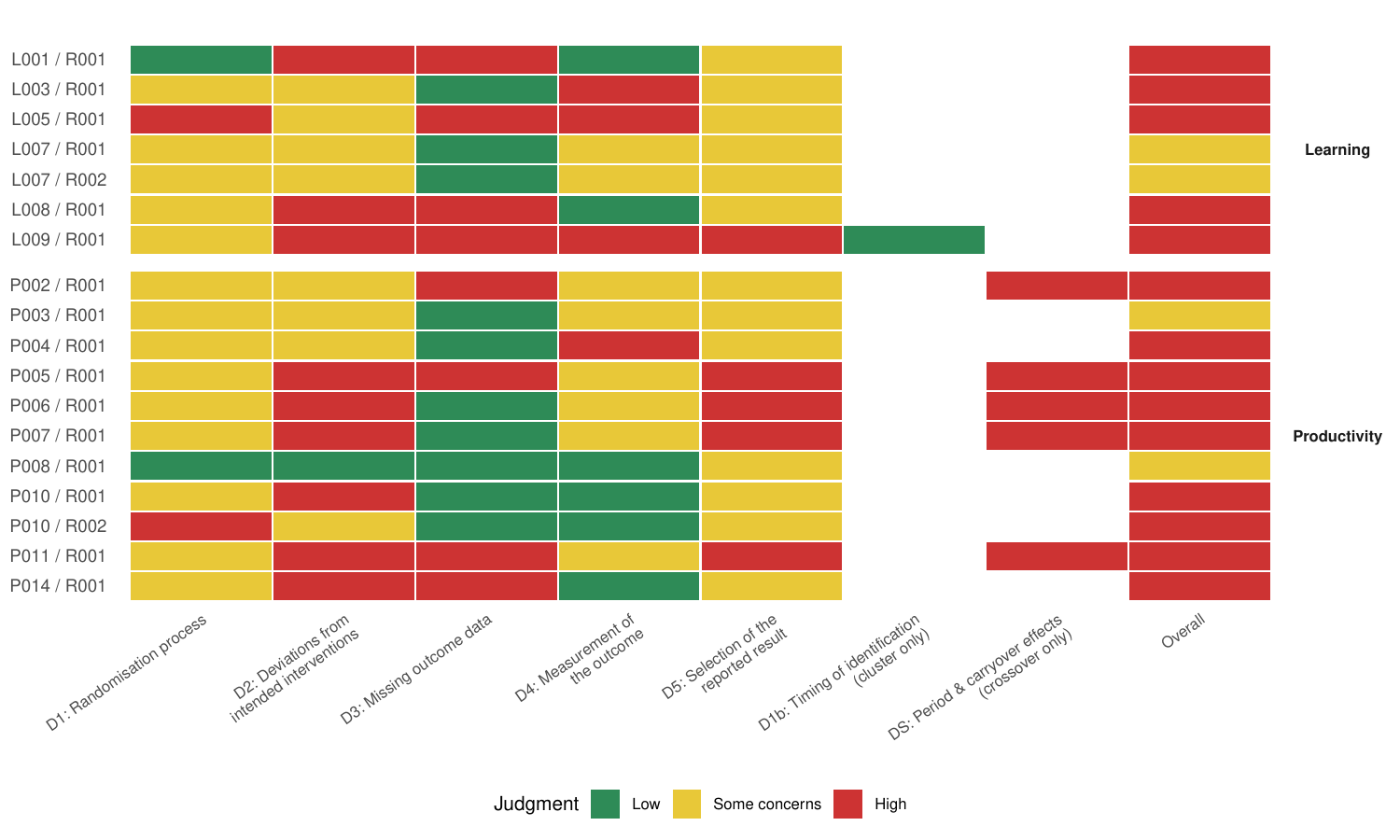}
    \caption{Risk-of-bias assessment of included studies using the revised Cochrane Risk of Bias tool (RoB~2). Each row represents a study--result pair, grouped by outcome domain (Learning and Productivity). Columns correspond to bias domains: D1:~Randomization process; D2:~Deviations from intended interventions; D3:~Missing outcome data; D4:~Measurement of the outcome; D5:~Selection of the reported result; D1b:~Timing of identification or recruitment of participants (cluster-randomized trials only); DS:~Period and carryover effects (crossover trials only). Judgments are color-coded as low risk of bias (green), some concerns (yellow), or high risk of bias (red).}
    \label{fig:rob2}
\end{figure}

\begin{figure}[htbp]
    \centering
    \includegraphics[width=\textwidth]{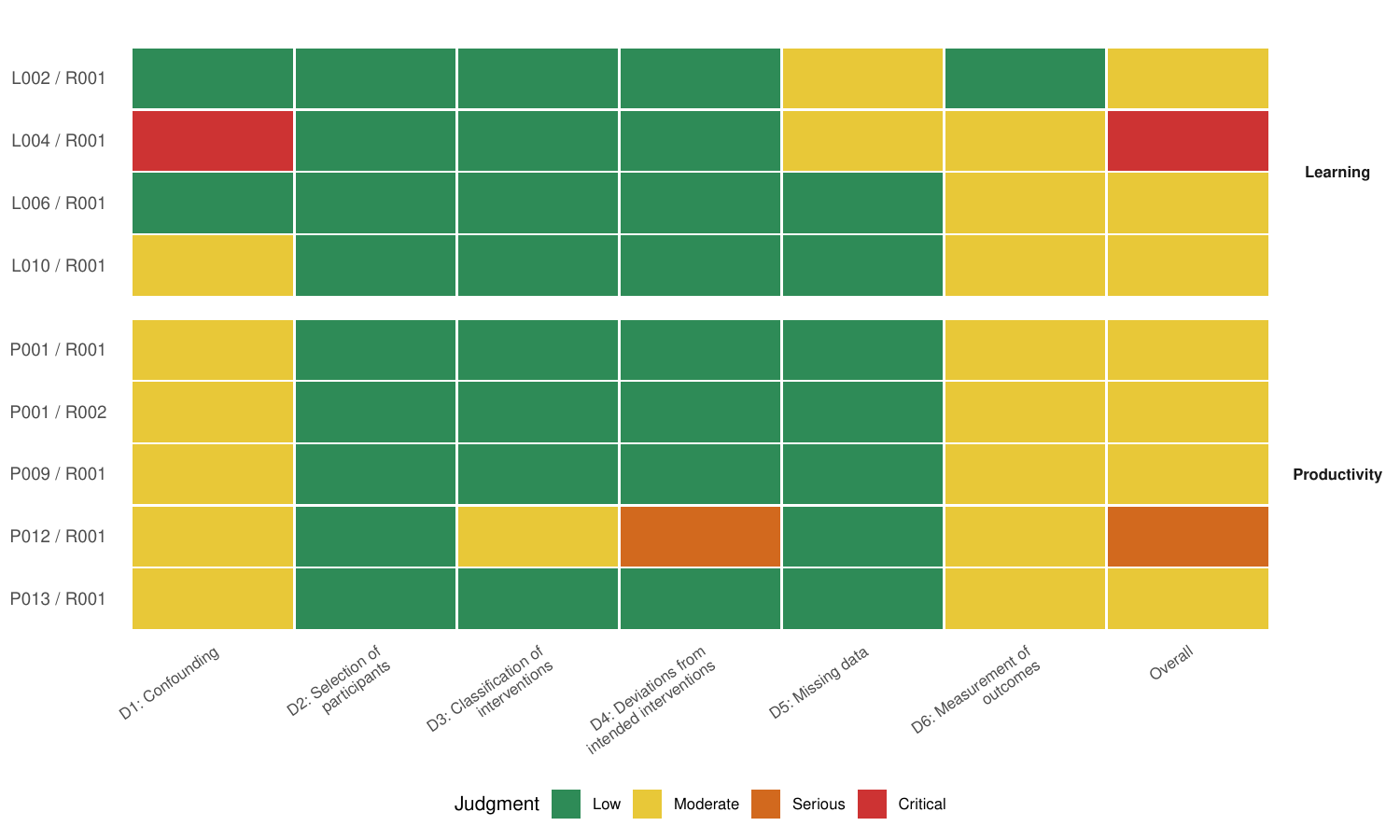}
    \caption{Risk-of-bias assessment of included non-randomized studies using the Risk Of Bias In Non-randomized Studies of Interventions (ROBINS-I) tool. Each row represents a study--result pair, grouped by outcome domain (Learning and Productivity). Columns correspond to bias domains: D1:~Confounding; D2:~Selection of participants; D3:~Classification of interventions; D4:~Deviations from intended interventions; D5:~Missing data; D6:~Measurement of outcomes. Judgments are color-coded as low risk of bias (green), moderate risk of bias (yellow), serious risk of bias (orange), or critical risk of bias (red).}
    \label{fig:robins-i}
\end{figure}

\newpage
\section{Effect Size Conversion Formulas}\label{appendix:es_conversions}

All effect sizes were computed as Cohen's $d$ and subsequently converted to 
Hedges' $g$ to correct for small-sample bias \cite{hedges_distribution_1981}. For each effect size, the corresponding sampling variance~$v$ was computed and used as the basis for inverse-variance weighting for the 
meta-analytic model \cite{borenstein_conversions2009}.
Five conversion paths were used, depending on the statistics 
reported in each primary study.
Throughout, $n_{\mathrm{T}}$ and $n_{\mathrm{C}}$ denote the treatment 
and control group sample sizes, and $z_{\alpha/2} = 1.96$ is the critical 
value for 95\% confidence intervals.
When only a standard error (SE) or confidence interval is reported, 
the standard deviation is recovered via 
$\mathit{SD} = \mathit{SE} \cdot \sqrt{n}$,
with $\mathit{SE} = (\mathrm{CI}_{\mathrm{upper}} - \mathrm{CI}_{\mathrm{lower}}) / (2 \times 1.96)$ 
where applicable.

\subsection{Within-Subject: Standardised Mean Change (Raw-Score SD)}
\label{appendix:smcr}

For single-group pretest--posttest designs, the effect size is the 
standardised mean change using the raw-score (pretest) standard deviation 
\cite{morrisCombiningEffectSize2002} 
\begin{equation}
  d_{\mathrm{SMCR}} 
  = \frac{\bar{X}_{\mathrm{post}} - \bar{X}_{\mathrm{pre}}}
         {\mathit{SD}_{\mathrm{pre}}}
\end{equation}
with sampling variance
\begin{equation}
  v = \frac{2(1-r)}{n}\;\frac{n-1}{n-3}\;
      \biggl(1 + \frac{n\,d_{\mathrm{SMCR}}^{2}}{2(1-r)}\biggr)
      - \frac{d_{\mathrm{SMCR}}^{2}}{\bigl[c(n-1)\bigr]^{2}},
\end{equation}
where $r$ is the pre--post correlation (set to $r = 0.5$ when unreported; 
\cite{morrisCombiningEffectSize2002}) and 
$c(\mathit{df}) = 1 - 3\,/\,(4\,\mathit{df} - 1)$ is the Hedges bias function.

\subsection{Between-Subject: Standardised Mean Difference}
\label{appendix:smd}

For independent-groups designs reporting group means and variation 
(SD, SE, or CI), the standardised mean difference 
\cite{borenstein_conversions2009} is computed:
\begin{equation}
  d = \frac{\bar{X}_{\mathrm{T}} - \bar{X}_{\mathrm{C}}}
           {\mathit{SD}_{\mathrm{pool}}},
  \qquad
  \mathit{SD}_{\mathrm{pool}} 
  = \sqrt{\frac{(n_{\mathrm{T}}-1)\,\mathit{SD}_{\mathrm{T}}^{2} 
              + (n_{\mathrm{C}}-1)\,\mathit{SD}_{\mathrm{C}}^{2}}
              {n_{\mathrm{T}} + n_{\mathrm{C}} - 2}}
\end{equation}
with sampling variance
\begin{equation}
  v = \frac{n_{\mathrm{T}} + n_{\mathrm{C}}}{n_{\mathrm{T}}\,n_{\mathrm{C}}} 
    + \frac{d^{2}}{2(n_{\mathrm{T}} + n_{\mathrm{C}})}
\end{equation}
%

\subsection{Pre-Post-Control Group Design: 
            \texorpdfstring{$d_{\mathrm{ppc2}}$}{dppc2} (Morris, 2008)}
\label{appendix:dppc2}

For studies with a pre-post-control group (PPC) design,
$d_{\mathrm{ppc2}}$ is used \cite{morrisEstimatingEffectSizes2008}.
This estimator standardizes the difference-in-differences by the 
\emph{pooled pre-test} standard deviation, which is unaffected by the 
treatment and therefore more consistent across studies. Where pre- and post-test instruments differed considerably in content or scoring (e.g., \cite{tangEnhancingProgrammingPerformance2025}), a post-test-only Hedges' $g$ was computed instead.

The pooled pre-test SD is:
\begin{equation}
  \mathit{SD}_{\mathrm{pool}} 
  = \sqrt{\frac{(n_{\mathrm{T}}-1)\,\mathit{SD}_{\mathrm{pre,T}}^{2} 
              + (n_{\mathrm{C}}-1)\,\mathit{SD}_{\mathrm{pre,C}}^{2}}
              {n_{\mathrm{T}} + n_{\mathrm{C}} - 2}}
\end{equation}

\paragraph{From raw pre/post means.}
When pretest and posttest means are available for both groups:
\begin{equation}
  d_{\mathrm{ppc2}} 
  = \frac{(\bar{X}_{\mathrm{post,T}} - \bar{X}_{\mathrm{pre,T}})
        - (\bar{X}_{\mathrm{post,C}} - \bar{X}_{\mathrm{pre,C}})}
         {\mathit{SD}_{\mathrm{pool}}}
\end{equation}

\paragraph{From a percentage treatment effect.}
When a study reports the treatment effect as a percentage of the 
control-group baseline ($\%_{\mathrm{eff}}$) together with the 
control-group baseline mean ($\bar{X}_{\mathrm{base,C}}$), 
the raw difference is first recovered:
\begin{equation}
  \Delta_{\mathrm{raw}} = \%_{\mathrm{eff}} \times \bar{X}_{\mathrm{base,C}},
  \qquad
  d_{\mathrm{ppc2}} = \frac{\Delta_{\mathrm{raw}}}{\mathit{SD}_{\mathrm{pool}}}
\end{equation}

\paragraph{Sampling variance (both variants).}
\begin{equation}
  v = 2(1-r)\Bigl(\frac{1}{n_{\mathrm{T}}} + \frac{1}{n_{\mathrm{C}}}\Bigr) 
    + \frac{d_{\mathrm{ppc2}}^{2}}{2(n_{\mathrm{T}} + n_{\mathrm{C}})}
\end{equation}
where $r$ is the pre--post correlation, set to $r = 0.5$ when unreported.

\subsection{Regression Coefficient to Standardized Effect}
\label{appendix:reg_beta}
For observational panel studies whose identification relies on controls absorbed within the model (e.g., unit and time fixed effects in TWFE/DiD designs), we computed effect sizes from $\hat\beta$ and the pooled standard deviation ($\mathit{SD}_{\mathrm{pooled}}$) where available.
\begin{equation}
  d = \frac{\hat\beta}{\mathit{SD}_{\mathrm{pooled}}},
  \qquad
  v = \biggl(\frac{\mathit{SE}_{\hat\beta}}
                   {\mathit{SD}_{\mathrm{pooled}}}\biggr)^{2}
\end{equation}

\textit{Note:} When separate pretest SDs for treatment and control groups were unavailable, we used $\mathit{SD}_{\mathrm{overall}}$ (pooled across all units and time periods). This yields a conservative effect size because $\mathit{SD}_{\mathrm{overall}}$ includes between-unit variance that the fixed effects have already absorbed from the numerator \cite{morrisEstimatingEffectSizes2008}.

\newpage
\section{Definitions of moderators}\label{appendix:definitions}

\begin{table}[H]
\centering
\small
\onehalfspacing
\setlength{\tabcolsep}{8pt}
\renewcommand{\arraystretch}{1.35}
\begin{tabularx}{\textwidth}{@{} p{3.8cm} X p{5.5cm} @{}}
\toprule
\textbf{Moderator} & \textbf{Description} & \textbf{Extracted values} \\
\midrule
Study setting
  & The context in which the study was conducted.
    \newline \textit{Levels:} Laboratory, Open-Source, Enterprise
  & Open-Source, Enterprise, Laboratory \\
GenAI interface
  & The specific IDE or interface provided to participants.
    \newline \textit{Levels:} GitHub Copilot, Other
  & GitHub Copilot, Cursor Pro, Codex, GILT, ClueBot, not stated \\
Programming language
  & Primary programming language of the task.
    \newline \textit{Levels:} Python, Other
  & Python, C++, Rust, JavaScript, Mixed \\
Participant level
  & Professional experience of participants. Developers contributing to open-source projects or working in an organizational context were
    classified as experienced, while the remaining studies involved student samples or mixed groups.
    \newline \textit{Levels:} Experienced, Students, Mixed
  & Experienced, Mixed, Students \\
Randomization
  & Whether treatment assignment was randomized.
    \newline \textit{Levels:} Randomized (e.g.\ RCT), Non-randomized (e.g.\ Quasi-Experiment)
  & Randomized, Non-randomized \\
\bottomrule
\end{tabularx}
\caption{Moderator variables for productivity outcomes included in the meta-regression analyses,
         with the specific values extracted from the included studies. Two pre-registered moderators were excluded: LLM base model was omitted because model specifications varied substantially and several studies did not report the exact model; coding task type was excluded as no meaningful categorization across studies was feasible.}
\label{tab:moderators_productivity}
\end{table}
 
\begin{table}[H]
\centering
\onehalfspacing
\small
\setlength{\tabcolsep}{8pt}
\renewcommand{\arraystretch}{1.35}
\begin{tabularx}{\textwidth}{@{} p{3.8cm} X p{5.5cm} @{}}
\toprule
\textbf{Moderator} & \textbf{Description} & \textbf{Extracted values} \\
\midrule
Exam environment
  & Whether GenAI tool use was permitted during the post-test assessment in the treatment condition.
    \newline \textit{Levels:} GenAI allowed, No GenAI
  & GenAI allowed, No GenAI allowed \\
GenAI interface
  & The specific IDE or interface provided to participants.
    \newline \textit{Levels:} ChatGPT, Other
  & ChatGPT, Codex, Custom interface \\
Programming language
  & Primary programming language of the task.
    \newline \textit{Levels:} Python, Other
  & Python, C++, C\#, Java \\
Study duration
  & Length of the AI-assisted learning intervention.
    \newline \textit{Levels:} ${<}$10 weeks, ${\geq}$10 weeks
  & Short-term ($<10$ weeks), Semester-long ($> 10$ weeks) \\
Participant educational level
  & Studies involved learners at varying educational stages, ranging from K-12 and high school students to
    university undergraduates and graduates, which we collapsed into two levels.
    \newline \textit{Levels:} University, Pre-University
  & Undergraduate, Graduate, Pre-university (K-12, High School) \\
Randomization
  & Whether treatment assignment was randomized.
    \newline \textit{Levels:} Randomized, Non-randomized
  & Randomized, Non-randomized \\
\bottomrule
\end{tabularx}
\caption{Moderator variables for learning outcomes included in the meta-regression analyses,
         with the specific values extracted from the included studies.}
\label{tab:moderators_learning}
\end{table}

\end{document}